\documentclass[a4paper,11pt,preprint]{article}
\pdfoutput=1

\usepackage{jcappub}
\usepackage{multirow}
\usepackage{graphicx}
\usepackage{dcolumn}
\usepackage{bm}
\usepackage{color}
\usepackage{epsfig}
\usepackage{hyperref}
\usepackage{natbib}
\usepackage{float}
\usepackage{array}
\usepackage{multirow}
\usepackage{ragged2e}
\usepackage{adjustbox}

\newcommand{\ra}[1]{\renewcommand{\arraystretch}{#1}}

\newcolumntype{L}{>{\displaystyle}l}
\newcolumntype{C}{>{\displaystyle}c}
\newcolumntype{R}{>{\displaystyle}r}

\definecolor{lightred}{rgb}{1,0.5,0.5}
\definecolor{lightgreen}{rgb}{0.5,1,0.5}
\definecolor{lightblue}{rgb}{0.5,0.5,1}
\definecolor{logblue}{rgb}{0,1,1}
\definecolor{lightcyan}{rgb}{0.5,0.75,0.75}
\definecolor{lightmagenta}{rgb}{0.75,0.5,0.75}
\definecolor{lightgreenred}{rgb}{0.75,0.75,0.5}
\definecolor{customgreen}{rgb}{0.494,1,0.502}
\definecolor{customblue}{rgb}{0,0,0.75}
\definecolor{gray}{rgb}{0.667,0.667,0.667}
\newcommand{\fn}[2]{\mathinner{#1\mathopen{\left(#2\right)}}} 

\title{Damping of an oscillating scalar field indirectly coupled to a thermal bath}

\author[a,b]{Erwin H. Tanin}
\author[b]{and Ewan D. Stewart}
\affiliation[a]{Center for Axion and Precision Physics Research, Institute for Basic Science (IBS),\\ Daejeon 34051, Republic of Korea}
\affiliation[b]{Department of Physics, Korea Advanced Institute of Science and Technology (KAIST),\\ Daejeon 34141, Republic of Korea}
\emailAdd{ehtanin@gmail.com}
\emailAdd{jcap@profstewart.org}
\date{\today}
\abstract{
The damping process of a homogeneous oscillating scalar field that indirectly interacts with a thermal bath through a mediator field is investigated over a wide range of model parameters. We consider two types of mediator fields, those that can decay to the thermal bath and those that are individually stable but pair annihilate. The former case has been extensively studied in the literature by treating the damping as a local effect after integrating out the assumed close-to-equilibrium mediator field. The same approach does not apply if the mediator field is stable and freezes out of equilibrium. To account for the latter case, we adopt a non-local description of damping that is only meaningful when we consider full half-oscillations of the field being damped. The damping rates of the oscillating scalar field and the corresponding heating rate of the thermal bath in all bulk parameter regions are calculated in both cases, corroborating previous results in the direct decay case. Using the obtained results, the time it takes for the amplitude of the scalar field to be substantially damped is estimated.  
}

\begin{document}
\maketitle

\section{Introduction}\label{s:introduction}
Supersymmetry is a well-motivated candidate for beyond the Standard Model physics. It solves the hierarchy problem, predicts the unification of gauge couplings at high energy, and provides potential dark matter candidates. With the recent measurement of the Higgs mass by the LHC being compatible with the MSSM upper bound, supersymmetry is now even more compelling than before \citep{Martin2016, Olive:2016xmw}. 

A generic feature of supersymmetric theories is the existence of a large number of directions in the scalar field space, called flat directions, along which the scalar potential vanishes up to supersymmetry-breaking and non-renormalizable terms \citep{Martin2016,Gherghetta1996}. Since it requires little energy to excite the flat directions, they can easily acquire a high degree of excitation in their low-momentum modes and yield various cosmological consequences, including: moduli problem \citep{Coughlan1983,Ellis1986,Banks1994,Carlos1993}, Affleck-Dine baryogenesis \citep{AD1985}, and thermal inflation \citep{Lyth1995,Lyth1996}. In some cases, a flat direction condensate may fragment into non-topological solitons \citep{Coleman1985,Kusenko1997b,Kasuya2002}, which may lead to even more varieties of cosmological consequences \citep{Enqvist2003, Enqvist2002b,Kusenko1997,Banerjee2000,Chiba2009}.

When a flat direction interacting with a thermal bath oscillates with a large amplitude, the fields it directly couples to gain large masses and so get driven out of thermal equilibrium, leaving the flat direction only indirectly coupled to the remaining thermal bath. The system would then seek higher entropy configurations by releasing the energy stored in the low-momentum modes of the flat direction to either the thermal bath or the high-momentum modes of the flat direction. Our objective in this paper is to assess the efficiency of such energy release. We consider a toy model with interaction configuration
\begin{equation*}
\phi\longleftrightarrow\psi\longleftrightarrow\chi
\end{equation*}
that is designed to mimic such a situation, where $\phi$ represents an oscillating scalar field parameterizing a flat direction, $\psi$ represents the fields $\phi$ couples to directly, and $\chi$ represents a thermal bath which, at tree level, couples to $\psi$ but not to $\phi$. For example, if $\phi$ represents the $H_u H_d$ flat direction in the MSSM, then $\psi$ represents the quark, electron, muon, tau, $W$, and $Z$ superfields while $\chi$ represents the neutrino, photon, and gluon superfields. Damping in systems with such \textit{two-stage} interaction configuration appear in many contexts and have been considered in \citep{instant,quasi,Drewes2013,Mukaida2012,Mukaida2013,Moroi2013,Mukaida2013b,Mukaida2014b,Moss2008,Gil2010}.

In the initial stages, long before $\phi$ thermalizes, there are two main damping effects on the $\phi$ oscillation. First, scatterings with particles of the mediator field $\psi$ excite the low-momentum modes of the $\phi$ field to its high-momentum modes, reducing the energy of the low-momentum modes. Second, the $\psi$-mediated interaction between $\phi$ and $\chi$ results in an entropically-favoured flow of energy from $\phi$ to $\chi$, an effect we dub \textit{mediated damping} for convenience. Our focus is on the latter process since it tends to be the dominant effect when the amplitude of the $\phi$ oscillation is large.

The simplest of the mediated damping processes considered in the literature is instant preheating \citep{instant}. To review the basic idea of instant preheating, let us first turn off the thermal effects due to the thermal bath $\chi$ by assuming that its temperature is extremely low. In such a case, the only role $\chi$ plays is as a decay/annihilation channel of $\psi$. As in broad parameteric resonance \citep{towards}, $\psi$ receives a periodically changing extra mass from the oscillating $\phi$ field, and whenever the mass of $\psi$ becomes small non-perturbative $\psi$ particle production takes place. However, this time, the created $\psi$ particles can decay/annihilate to $\chi$. It takes time for that to happen, and so the created $\psi$ particles begin to gain mass again, extracting energy from $\phi$ in the process. By the time the $\psi$ particles decay/annihilate to $\chi$, they are already significantly more massive than when they were created, i.e.\ some amount of energy has been extracted from $\phi$ and dumped to $\chi$.

The situation becomes more complicated in the presence of a high-temperature thermal bath $\chi$. At any given time, the interaction between $\psi$ and $\chi$ tends to bring $\psi$ to thermal equilibrium with $\chi$. However, since the mass of $\psi$ is always changing, so is its thermal equilibrium abundance. As it takes finite time for $\psi$ to equilibrate with $\chi$, we are faced with a non-equilibrium problem where the abundance of $\psi$ is perpetually seeking its elusive instantaneous thermal equilibrium value. The non-equilibrium nature of the problem makes it challenging to study analytically.

Starting from the pioneering works in \citep{Hosoya1984,Morikawa1984}, analytical studies on mediated damping involving a high temperature thermal bath were usually done by assuming that there is a separation of timescales between the slow dynamics of the $\phi$ field and the fast dynamics of the remaining fields. In that case, the $\psi$ field, assumed to be always in quasi thermal equilibrium with the thermal bath, can be integrated out to give us the coarse-grained equation of motion of the $\phi$ field, where the effect of the $\psi$ field manifests as a local friction term that is proportional to $\dot{\phi}$. In this local limit, the mediated damping effect has been given several interpretations: as a transport phenomenon analogous to electrical or thermal conductivity \citep{Hosoya1984}, as an effect of particle-production due to time-dependent mass of the mediator field (even when the mass evolution is adiabatic) \citep{Morikawa1984,Moss2008}, and as an effective three-point decay process \citep{Yokoyama:2004pf, Mukaida2013}.

To allow the mediator field $\psi$ to remain close to thermal equilibrium, previous studies were limited to cases where the $\psi$ particles can decay to the thermal bath. However, scenarios where the $\psi$ particles are protected by some symmetry, and hence stable, also occur naturally. In those cases, we expect the same mediated damping effect to be present, but the problem is that the stability of $\psi$ implies that it cannot remain close to thermal equilibrium at all times due to the freeze out phenomenon. This means that the usual local approach does not apply. 

From a non-local perspective, the mediated damping effect can be understood as a process similar to instant preheating, where $\psi$ particles are created when they are light, absorb energy from $\phi$ as their masses increase, and later decay/annihilate to the thermal bath $\chi$. Unlike in instant preheating, the thermal bath $\chi$ does not merely serve as an energy dumping site. It could also do the job of thermally exciting $\psi$ when it is light. Alternatively, one could view the damping from $\phi$'s perspective. In this view, the damping emerges from the backreaction of $\psi$ in the form of an extra potential for $\phi$ that is less steep in the descending part and steeper in the ascending part of the $\phi$ oscillation.
 
In this paper, we tackle the problem using an elementary approach and treat the $\phi$ damping as a non-local effect that is only meaningful when we consider full half-oscillations of the $\phi$ field. This approach allows us to estimate the damping rate even in situations where the $\psi$ field is far from thermal equilibrium. We consider the case where single-particle decay of $\psi$ is allowed for a consistency check with previous results \citep{Mukaida2012,Mukaida2013} and apply the same method to the case where $\psi$ is stable.

This paper is organized as follows. In Section~\ref{s:the model}, we introduce our model and lay out the key assumptions we use. In Section~\ref{s:mediated damping process}, we explain the basic concept of mediated damping. In Section~\ref{s:dynamics of psi}, we study the dynamics of $\psi$, which determines all energy flow in the system. In Section~\ref{s:damping of phi in one half-oscillation}, we calculate the amount of $\phi$ energy damped due to the mediated damping in one half-oscillation of $\phi$ in a way that is independent of the details of the $\phi$ oscillation. In Section~\ref{s:effective potential of phi}, we study the different types of effective potential underlying the oscillation of $\phi$ due the backreaction from $\psi$ and their regimes of applicability. In Section~\ref{s:results}, we analyze the long-term behaviour of the system over many oscillations of $\phi$. In Section~\ref{s:summary}, we provide a summary of the preceding results. Readers who are not interested in the technical details may skip to Section~\ref{s:summary} after reading Section~\ref{s:the model} and \ref{s:mediated damping process}. Finally, Section~\ref{s:discussion} and \ref{s:conclusion} are devoted to discussion and conclusion.

\section{The Model}\label{s:the model}
In the actual supersymmetric flat direction damping, the fields involved are superfields, but here we simplify them to just two real scalar fields, $\phi$ and $\psi$, and a thermal bath $\chi$ of either real scalar or spinor fields. The Lagrangian of our system is
\begin{align}
\mathcal{L}=&\frac{1}{2}\left(\partial_\mu\phi\right)^2+\frac{1}{2}\left(\partial_\mu\psi\right)^2-\frac{1}{2}m_\phi^2\phi^2-\frac{1}{2}m_\psi^2\psi^2-\frac{1}{2}\lambda^2\phi^2\psi^2-V_{\psi\chi}[\psi,\chi]+\mathcal{L}_{\text{bath}}[\chi]
\end{align}
where $\lambda$ is a Yukawa coupling constant, whose value may range over many orders of magnitude. In the context of supersymmetry, the bare masses $m_\phi$ and $m_\psi$ are expected to be of the order of the supersymmetry breaking scale. Hence, we assume
\begin{align}
m_\phi\sim m_\psi \label{susybreaking}
\end{align}
Here, the thermal bath $\chi$ has $g_*>>1$ degrees of freedom, and, for simplicity, we neglect its thermal equilibration time, i.e.\ we assume that $\chi$ is always in a thermal state. With the latter assumption, the details of $\mathcal{L}_{\text{bath}}[\chi]$, which determines the dynamics of $\chi$, is not important since the relevant properties of $\chi$ can always be extracted from its temperature $T$.

We consider separately the case where the single-particle decay $\psi\rightarrow\chi\chi$ is allowed and the case where $\psi$ particles are stable and have to rely on the annihilation process $\psi\psi\rightarrow\chi\chi$ to convert to $\chi$ particles. In the former case, which we will refer to as the \textit{decay case}, we take $\chi$ to be a collection of identical spinor fields $\{\chi_i\}$ which interact with $\psi$ through Yukawa interactions of the form
\begin{align}
V_{\psi\chi}^{\text{D}}[\psi,\chi]=\sum_{i=1}^{g_*/2}\kappa_{\text{D}}\psi\bar{\chi}_i\chi_i\label{V_dec}
\end{align}
In the latter case, which we will refer to as the \textit{annihilation case}, we take $\chi$ to be a collection of identical real scalar fields $\{\tilde{\chi}_i\}$ which interact with $\psi$ through Yukawa interactions of the form
\begin{align}
V_{\psi\chi}^{\text{A}}[\psi,\chi]=\sum_{i=1}^{g_*}\frac{1}{2}\kappa_{\text{A}}^2\psi^2\tilde{\chi}_i^2\label{V_ann}
\end{align}
In the example of the $H_u H_d$ flat direction we gave previously, the unstable $\psi$ field represents the $Z$ superfield and the stable $\psi$ field represents the quark, electron, muon, tau, and $W$ superfields\footnote{In this example, there is the added complication that the stable $\psi$ fields can decay into one another.}. If a $\psi$ field can both decay and annihilate to $\chi$, the decay effect would be dominant. On the other hand, if $\phi$ couples to both stable and unstable $\psi$ fields we expect the $\phi$ damping effect to resemble that due to the stable $\psi$ fields since, due to freeze out, they are more abundant in general and so tend to give a stronger backreaction to $\phi$.

The $\phi$ field models a field that is highly excited only in its low-momentum modes. For simplicity and solubility, we will treat $\phi$ as a homogeneous classical field. The amplitudes of the initially barely excited inhomogeneous modes of $\phi$ will build up over time, but this process tends to be slow. Presumably, by the time the inhomogeneous modes of $\phi$ are excited highly enough to give significant effects, the goal of damping the homogeneous mode of $\phi$ has been achieved. This is at least the case if the inhomogeneous modes of $\phi$ are excited solely by $\phi$-$\psi$ scattering. Other possible sources of spatial fluctuations of $\phi$ are briefly discussed in the Discussion section.

\section{Mediated Damping Process} \label{s:mediated damping process}
Consider the dynamics of the system during one half-oscillation of $\phi$ where $\phi$ goes from a large negative value to a large positive value. When $\phi$ descends down the potential, the mass of $\psi$ decreases, and hence the thermal equilibrium abundance of $\psi$ rises. As this happens, the actual abundance of $\psi$ lags below and chases the equilibrium value. After $\phi$ passes $\phi=0$, the mass of $\psi$ increases and the equilibrium abundance decreases. At some point, the $\psi$ abundance catches up with the equilibrium value, but then it overshoots and lags above the equilibrium value until the end of the half-oscillation. 

An important observation is that while the thermal equilibrium abundance is symmetrical in the descending and ascending part of the oscillation of $\phi$, $\psi$ is more abundant in the ascent because of the way it lags behind the equilibrium abundance. Hence, the extra potential felt by $\phi$, which is roughly speaking proportional to the abundance of $\psi$, is asymmetric, less steep during the descent and steeper during the ascent. As a result, the net effect after one half-oscillation of $\phi$ is a decrease in the amplitude of $\phi$. 

So far, we have disregarded non-adiabatic $\psi$ particle production. The occurrence of non-adiabatic $\psi$ particle production will only enhance the asymmetry, since it increases the abundance of $\psi$ right at the transition from the descending part to the ascending part of the oscillation of $\phi$, thus making the potential in the ascent even steeper compared to that in the descent.

In the previous description, the mediated damping process is understood from $\phi$'s perspective. The same damping effect can also be seen from $\psi$'s point of view, one which we will mainly use for our calculations, in a way analogous to instant preheating. The $\psi$ field is prone to excitation when its mass is small, at which point exciting $\psi$ is energetically cheap. The $\phi$ field can then exchange energy with the created $\psi$ particles by modifying the latter's mass\footnote{Strictly speaking, the energy contained in the extra mass of $\psi$ due to its interaction with $\phi$ does not belong to $\phi$ or $\psi$ alone. It belongs to the interaction of the two fields. The same goes for the $\psi$-$\chi$ interaction. For convenience, we group any interaction energy into the $\psi$ sector here.}. Later, some of the created $\psi$ particles will decay/annihilate to the thermal bath when they are relatively heavier. The net effect after such decay/annihilation is energy being taken from $\phi$ and dumped to $\chi$.

In general, it makes sense to talk about the mediated damping of $\phi$ only if we consider a whole half-oscillation of $\phi$. However, in the limit where $\psi$ closely tracks its equilibrium abundance, the damping can be seen locally as due to a slight phase delay between the oscillation of $\phi$ and the backreaction from $\psi$ to $\phi$. In that case, the backreaction due to $\psi$ manifests in the equation of motion of $\phi$ as a traditional friction term that is proportional to $\dot{\phi}$.

During each half-oscillation of $\phi$, $\phi$ and $\psi$ transfer energy back and forth. At the end of a half-oscillation, a large amount of energy of $\phi$ may end up in the form of the mass energy of $\psi$ particles. This contributes to decreasing the amplitude of $\phi$ along with mediated damping and other damping effects. However, in the next quarter oscillation part of the energy taken from $\phi$ will be returned back to $\phi$ in the form of kinetic energy when the mass of $\psi$ becomes small again. Since such energy transfer is time-reversible, we refer to such amplitude decrease as trapping rather than damping. More specifically, we define trapping as the decrease in the amplitude of $\phi$ due to the symmetric part of the effective potential of $\phi$. What we refer to as damping is when energy is taken out of $\phi$ irreversibly. For instance, mediated damping occurs when energy is taken from $\phi$ and then dumped to $\chi$. Such energy transfer is irreversible because the relaxation of $\chi$ to thermal equilibrium after receiving the energy is an irreversible process.

We now proceed with our calculations, as outlined at the end of Section~\ref{s:introduction}. Readers who are not interested in the technical details may jump to Section~\ref{s:summary}.

\section{Dynamics of \boldmath $\psi$}\label{s:dynamics of psi}
All flow of energy in our system involves $\psi$, so to understand the energy flow in the system it is sufficient to study $\psi$'s energy input and output. Hence, we devote this section to study the dynamics of $\psi$.

\subsection{Assumptions}\label{ss:assumptions}
As described in Section~\ref{s:mediated damping process}, the $\phi$ field will oscillate asymmetrically in its ascent and descent. The exact solution of $\phi$ can be decomposed into its symmetric part $\phi_{\text{s}}$ and asymmetric part $\phi_{\text{a}}$
\begin{align}
\phi=\phi_{\text{s}}+\phi_{\text{a}}
\end{align}
For the cases of our interest, it can be shown that the asymmetric part is much smaller compared to the symmetric part and can be treated as a perturbation. The oscillating $\phi$ field induces an extra time-dependent mass on $\psi$. We approximate this extra mass using the symmetric part $\phi_{\text{s}}$. Furthermore, $\psi$ also receives a thermal mass \citep{Dolan1974}
\begin{align}
m_{\psi, \text{th}}^2=
\frac{1}{12}g_* \kappa_{\text{D/A}}^2T^2
\end{align}
due to its interaction with the thermal bath $\chi$. Altogether, the effective mass $M_\psi$ of $\psi$ can be written as
\begin{align}
M_\psi^2(t)=m_\psi^2+m_{\psi, \text{th}}^2+\lambda^2\phi_s^2(t)\label{mass}
\end{align}

We regard the temperature $T$ as a constant over a half-oscillation of $\phi$ and later justify it a posteriori by showing that the maximum amount of energy taken from or given to $\chi$ during one half-oscillation of $\chi$ is always much less than the energy contained in $\chi$. This might not be true when the temperature $T$ is small. Nevertheless, in that case the damping is dominated by non-adiabatic particle production effect that does not depend on $T$.

To focus on the mediated damping effect, we assume the following throughout the paper
\begin{align}
m_\phi,m_\psi&<<\lambda A \label{amplitude}\\
\lambda^4&<<\kappa\label{lambdag}\\
m_\phi&<M_\psi\label{mumphi}
\end{align}
where $A$ is the amplitude of oscillation of $\phi$ and we have made use of a new constant $\kappa$, defined as
\begin{align}
\kappa\equiv\begin{cases}
\displaystyle g_*\kappa_{\text{D}}^2 & \text{decay}\\[2ex]
\displaystyle g_*\kappa_{\text{A}}^4 & \text{annihilation}
\end{cases} \label{effectivecoupling}
\end{align}
$\kappa$ can be understood as the effective coupling strength between $\psi$ and $\chi$ collectively. For simplicity, we take
\begin{align}
\kappa\sim 1
\end{align}
but still perturbative. Eq.~\eqref{amplitude} indirectly implies that $M_\psi$ evolves adiabatically during most part of $\phi$'s oscillation, allowing us to perform analysis in particle picture. Eq.~\eqref{lambdag} allows us to neglect the annihilation of $\psi$ particles back to $\phi$ when evaluating the dynamics of $\psi$\footnote{Note that the annihilation of $\psi$ particles to $\phi$ will not be severely Bose-enhanced. Since the energy of a typical $\psi$ particle is always much greater than the mass of $\phi$, if the $\psi$ particles are to decay/annihilate to $\phi$ they will not fill the highly excited low-momentum modes of $\phi$.}. Eq.~\eqref{mumphi} ensures that the perturbative annihilation from $\phi$ to $\psi$ is kinematically blocked. The last assumption is not a strong one. It can be shown that if it does not hold initially it takes roughly a half-oscillation of $\phi$ for the temperature of the thermal bath to increase to the point where the thermal mass $m_{\psi, \text{th}}$ of $\psi$ is much bigger than $m_\phi$.

During most part of $\phi$'s oscillation where $M_\psi$ evolves adiabatically the $\psi$ field can be regarded as a gas of particles with number density evolving roughly according to the integrated Boltzmann equation \citep{Kolb:1990vq}
\begin{align}
\dot{n}_\psi=\Gamma_{\psi\rightarrow\chi\chi}\left(n_\psi^{\text{eq}}-n_\psi\right)\label{boltzmannd}
\end{align}
for the decay case and
\begin{align}
\dot{n}_\psi=\left<\sigma_{\psi\psi\rightarrow\chi\chi}v_{\text{rel}}\right>\left[\left(n_\psi^{\text{eq}}\right)^2-n_\psi^2\right]\label{boltzmanna}
\end{align}
for the annihilation case. $n_\psi^{\text{eq}}$ is the instantaneous thermal equilibrium abundance of $\psi$, $\Gamma_{\psi\rightarrow\chi\chi}$ is the $\psi\rightarrow\chi\chi$ single-particle decay rate, and $\left<\sigma_{\psi\psi\rightarrow\chi\chi}v_{\text{rel}}\right>$ is the thermally averaged cross-section times relative velocity for the $\psi\psi\rightarrow\chi\chi$ process. In the cases of our interest, where non-relativistic approximation is accurate up to $O(1)$ factors, $\Gamma_{\psi\rightarrow\chi\chi}$ for the decay case and $\left<\sigma_{\psi\psi\rightarrow\chi\chi}v_{\text{rel}}\right>$ for the annihilation case can be approximated as (Appendix \ref{s:cross section times relative velocity}, ref. \citep{Peskin1995})
\begin{align}
\Gamma_{\psi\rightarrow\chi\chi}&\sim \kappa M_\psi\label{decayrate}\\
\left<\sigma_{\psi\psi\rightarrow\chi\chi}v_{\text{rel}}\right>&\sim  \frac{\kappa}{ M_\psi^2}\label{crosssection}
\end{align}
The integrated Boltzmann equations \eqref{boltzmannd},\eqref{boltzmanna} are applicable if the kinetic equilibration rate of $\psi$ is much faster than its chemical equilibration rate and Bose-enhancement effects are negligible. Strictly speaking, the two conditions are not always satisfied in our system, but, since we are not concerned about $O(1)$ factors, the simplified Boltzmann equations are sufficient for our purposes.

\subsection{\boldmath $\psi$-particle Production}\label{ss:psi particle production}
When $|\phi_s|$, and hence $M_\psi$, is small, it becomes easy to excite $\psi$ and rapid $\psi$-particle production may take place. This can happen through thermal excitation and/or violation of the adiabaticity in the evolution of $M_\psi$. If both effects are present concurrently, the behaviour of the system becomes complex and difficult to solve. Nevertheless, as we will show, there are soluble extreme cases where one of the effects is negligible compared to the other. Solving these extreme cases would give us some ideas on how the system generally behaves. Hence, we start by considering the two effects exclusively.

\subsubsection{Thermal Excitation}\label{sss:thermal excitation}
Thermal excitation of $\psi$ has considerable effect when $\psi$ can be thermally excited by the thermal bath without Boltzmann suppression. It happens when
\begin{align}
M_\psi\lesssim T\label{th}
\end{align}
We will subsequently refer to this mass regime as the \textit{thermal regime}. In the thermal regime, the  Boltzmann equation \eqref{boltzmannd},\eqref{boltzmanna} for both the decay and annihilation case can be written as
\begin{align}
\dot{n}_\psi\approx \Gamma_{\text{th}}\left(n_\psi^{\text{eq}}-n_\psi\right) \label{simboltz}
\end{align} 
where
\begin{align}
\Gamma_{\text{th}}\sim \kappa T
\end{align}
is the thermalization rate of $\psi$. In this regime, $\psi$ can be thermally excited at most up to $n_\psi^{\text{eq}}\sim T^3$,
at which point $\psi$ is in thermal equilibrium with the thermal bath. In general, however, the system may not spend enough time in the thermal regime for $\psi$ to achieve full thermal equilibrium. Thus, more generally, $\psi$ can be thermally excited up to
\begin{align}
n_\psi^{\text{th}}&\sim T^3\min\left(1,\Gamma_{\text{th}}\Delta t_{\text{th}}\right) \label{thexcitation}
\end{align}
where $\Delta t_{\text{th}}\sim T/|\dot{M}_\psi|$ is the duration the system spends in the thermal regime. Note that, it does not mean that a particle density $n_\psi^{\text{th}}$ will remain after the system escapes from the thermal regime. The abundance of $\psi$ may continue to track its equilibrium value and get Boltzmann-suppressed. The above estimation serves as a measure for the strength of thermal excitation near the origin.

\subsubsection{Non-adiabatic Particle Production}\label{sss:non-adiabatic particle production}
When $M_\psi$ changes non-adiabatically, non-perturbative vacuum production of $\psi$ particles takes place. The condition for non-adiabaticity in the evolution of $M_\psi$ can be written as \citep{towards,beauty}
\begin{align}
M_\psi\lesssim |\dot{M}_\psi|^{1/2} \label{NA}
\end{align}
We will subsequently refer to this mass-regime as the \textit{non-adiabatic regime}. During the bulk of one half-oscillation of $\phi$, the order of magnitude of $|\dot{M}_\psi|^{1/2}$ typically stays the same. Thus, the evolution of $M_\psi$ is maximally non-adiabatic when $M_\psi$ is small, and the non-adiabatic regime exists if the minimum effective mass of $\psi$ is smaller than or of the same order with $|\dot{M}_\psi|^{1/2}$. After a passage of $\phi$ through the non-adiabatic regime, $\psi$ particles with number density
\begin{align}
n_\psi^{\text{NA}}\sim |\dot{M}_\psi|^{3/2} \label{NAparticle}
\end{align}
and typical momentum $k_{\text{NA}}\sim |\dot{M}_\psi|^{1/2}$ are produced \citep{towards,beauty}. This is, however, only true if the interaction between $\psi$ and $\chi$ does not disturb the particle production process, which we assume to be the case. As done in Ref. \citep{Mukaida2013}, we can show that thermal effects do not sabotage the non-adiabatic particle production using particle-picture estimations. Eq.~\eqref{NAparticle} gives the number of $\psi$ particle produced when there is no pre-existing $\psi$ particles. If a given $k$-mode of $\psi$ is already occupied, the number of created $\psi$ particles belonging to that mode will be Bose-enhanced.

\subsubsection{Thermal Effect vs Non-adiabatic Effect}\label{sss: thermal effect vs non-adiabatic effect}
The nature of $\psi$-particle production process near the origin is largely determined by the ratio between two important scales of our system, the temperature $T$ of the thermal bath and a scale $|\dot{M}_\psi|^{1/2}$ representing the rate of change of $\psi$'s effective mass. First, the non-adiabatic regime exists if
$\min\left(M_\psi\right)\approx m_{\psi, \text{th}}\lesssim|\dot{M}_\psi|^{1/2}$, which translates to
\begin{align}
\frac{T}{|\dot{M}_\psi|^{1/2}}\lesssim \sqrt{\frac{12}{g_*}}\frac{1}{\kappa_{\text{D/A}}}
\end{align}
If the non-adiabatic regime exists, the relative size between the thermal and non-adiabatic mass-regime is given by
\begin{align}
\frac{\text{size of thermal mass-regime}}{\text{size of non-adiabatic mass-regime}}\sim\frac{T}{|\dot{M}_\psi|^{1/2}}\label{relativesize}
\end{align}
Furthermore, the expression 
\begin{align}
\Gamma_{\text{th}}\Delta t_{\text{th}}\sim \frac{\kappa T^2}{|\dot{M}_\psi|}
\end{align}
in Eq.~\eqref{thexcitation} determines how closely the $\psi$ abundance tracks its thermal equilibrium abundance in the thermal regime. The larger (smaller) $\kappa T^2/|\dot{M}_\psi|$ is the slower (faster) the mass of $\psi$ varies compared to the thermalization timescale, and hence the easier (harder) it is for $\psi$ to track its thermal equilibrium state. Moreover, the ratio between the number density of $\psi$ particles created through thermal and non-adiabatic mass evolution effect, if both occur in one passing of $\phi$ near the origin without interfering with each other, is given by
\begin{align}
\frac{n_\psi^{\text{th}}}{n_\psi^{\text{NA}}}\sim \frac{T^3}{|\dot{M}_\psi|^{3/2}}\min\left(1, \frac{\kappa T^2}{|\dot{M}_\psi|}\right)
\end{align}

For simplicity, we will blur the quantitative differences between the thermal mass $m_{\psi,\text{th}}$, thermalization rate $\Gamma_{\text{th}}$, and temperature $T$
\begin{align}
m_{\psi,\text{th}}\sim \Gamma_{\text{th}}\sim T
\end{align}
The ratio between $T$ and $|\dot{M}_\psi|^{1/2}$ then tells us about: whether or not the non-adiabatic regime exist, the relative size between the non-adiabatic and thermal mass-regime, how closely the abundance of $\psi$ tracks its thermal equilibrium value, and the relative strength between non-adiabatic and thermal particle production.

\section{Damping of \boldmath $\phi$ in One Half-oscillation}\label{s:damping of phi in one half-oscillation}
In all the cases we consider, the ultra-relativistic modes of $\psi$ never get significantly excited, which means non-relativistic analyses are accurate up to factors of order one. Therefore, the energy density $\rho_\psi$ of $\psi$ outside the non-adiabatic regime can be written in terms of $\psi$'s number density $n_\psi$ and mass $M_\psi$ as
\begin{align}
\rho_\psi\sim n_\psi M_\psi
\end{align}
and its rate of change is
\begin{align}
\dot{\rho}_{\psi}\sim n_\psi\dot{M}_\psi+\dot{n}_\psi M_\psi \label{energyrate}
\end{align}
To get to know the direction of energy flow, from or to $\phi$ or $\chi$, that yields the above change in the energy density, one needs to understand the nature of changes in $M_\psi$ and $n_\psi$. $M_\psi$ is given by Eq.~\eqref{mass} and $n_\psi$ evolves according to Eq.~\eqref{boltzmannd} and Eq.~\eqref{boltzmanna}. Since we are assuming that the temperature of the thermal bath is unchanging over one half-oscillation of $\phi$, only $\psi$'s interaction with $\phi$ can change the mass of $\psi$ significantly. Furthermore, outside the non-adiabatic regime, only $\psi$'s interaction with $\chi$ can change the number density of $\psi$ significantly. So, speaking about energy transfer, the first and second term of Eq.~\eqref{energyrate} are essentially $\psi$'s only mean of communicating with $\phi$ and with $\chi$ respectively. 

The energy per unit volume $\Delta \rho_{\phi\rightarrow\psi}$ transferred from $\phi$ to $\psi$ in a half-oscillation of $\phi$ can thus be calculated by integrating the first term of Eq.~\eqref{energyrate}
\begin{align}
\Delta \rho_{\phi\rightarrow\psi}\sim \Delta \rho_{\phi\rightarrow\psi}^{\text{NA}}+\int_{\text{half-osc}}n_\psi \dot{M}_\psi  dt\label{damp}
\end{align}
where $\Delta \rho_{\phi\rightarrow\psi}^{\text{NA}}$ is the contribution from the non-adiabatic regime\footnote{The energy flow in the non-adiabatic regime, where the particle picture does not apply, will be discussed separately where it is needed.}. Similarly, the energy per unit volume $\Delta\rho_{\psi\rightarrow\chi}$ transferred from $\psi$ to $\chi$ is determined by the negative of the integral of the second term of Eq.~\eqref{energyrate}
\begin{align}
\Delta\rho_{\psi\rightarrow\chi}\sim  \Delta\rho_{\psi\rightarrow\chi}^{\text{NA}}-\int_{\text{half-osc}}\dot{n}_\psi M_\psi dt\label{temperature}
\end{align}

As explained in Section~\ref{s:mediated damping process}, $\Delta\rho_{\phi\rightarrow\psi}$ does not determine the $\phi$ damping because part of the energy transferred from $\phi$ to $\psi$ after one half-oscillation will be returned to $\phi$ when the mass of $\psi$ becomes small again in the next quarter of oscillation. It is $\Delta\rho_{\psi\rightarrow\chi}$ that determines the irreversible energy leak to $\chi$, i.e.\ the damped energy. When the system is in a steady state with $\Delta \rho_\psi=\Delta \rho_{\phi\rightarrow\psi}-\Delta\rho_{\psi\rightarrow\chi}=0$ the two are equal. In that case, sometimes it is easier to calculate $\Delta\rho_{\phi\rightarrow\psi}$ instead. 

In this section, we will evaluate Eq.~\eqref{temperature} in three cases characterized by the size of the thermal mass-regime $T$ as compared to the size of non-adiabatic mass-regime $|\dot{M}_\psi|^{1/2}$ and $\lambda A$, the amplitude of the modulation of $M_\psi(t)$. Before actually doing that, we make the following simplifications. It can be shown case-by-case that the orders of magnitude of $\dot{\phi}_s$ and $|\dot{M}_\psi|$ do not change within the regime where the damping mainly takes place. Since none of the effects we are considering depends on the derivatives of these variables and we do not pay attention to $O(1)$ factors, we can assume that both $\dot{\phi}_s$ and $|\dot{M}_\psi|$ are constants in the bulk of an oscillation of $\phi$. To further simplify our calculations, we can take
\begin{align}
\phi_s(t)=|\dot{\phi}_s|t\label{unperturbed}
\end{align}
where the time $t$ has been defined in such a way that $\phi_s(t=0)=0$. We will also let the time run from $t=-t_{\text{osc}}$ to $t=t_{\text{osc}}$ for one complete half-oscillation of $\phi$ to avoid writing factors such as $2$ or $4$. With the above simplifications, the time regime within which $\psi$ can be thermally excited without Boltzmann suppression is
\begin{align}
|t|\lesssim t_{\text{th}}=\frac{T}{|\dot{M}_\psi|}
\end{align}
and the time regime in which the mass of $\psi$ changes non-adiabatically is
\begin{align}
|t|\lesssim t_{\text{NA}}=\frac{1}{|\dot{M}_\psi|^{1/2}} \label{timena}
\end{align}

\subsection{Low Temperature Regime: \boldmath $T<<|\dot{M}_\psi|^{1/2}$}\label{ss:low temperature regime}
In this case, thermal excitation of $\psi$ can be neglected and the non-adiabatic regime is much larger than the thermal regime. The latter implies that the thermal equilibrium abundance of $\psi$ is Boltzmann suppressed outside the non-adiabatic regime, and the evolution of $n_\psi$ outside the non-adiabatic regime is described by Boltzmann Eqs.~\eqref{boltzmannd} and \eqref{boltzmanna} with $n_\psi^{\text{eq}}=0$
\begin{align}
\dot{n}_\psi\sim \begin{cases}
\displaystyle  \hfil -M_\psi n_\psi     & \text{ decay} \\[2ex]
\displaystyle \hfil-\frac{n_\psi^2}{M_\psi^2}  & \text{ annihilation} \label{ssimboltz}
  \end{cases}
\end{align}

\subsubsection{Decay Case}\label{sss1:decay case}
\begin{figure}[h]
\centering
\includegraphics[scale=0.45]{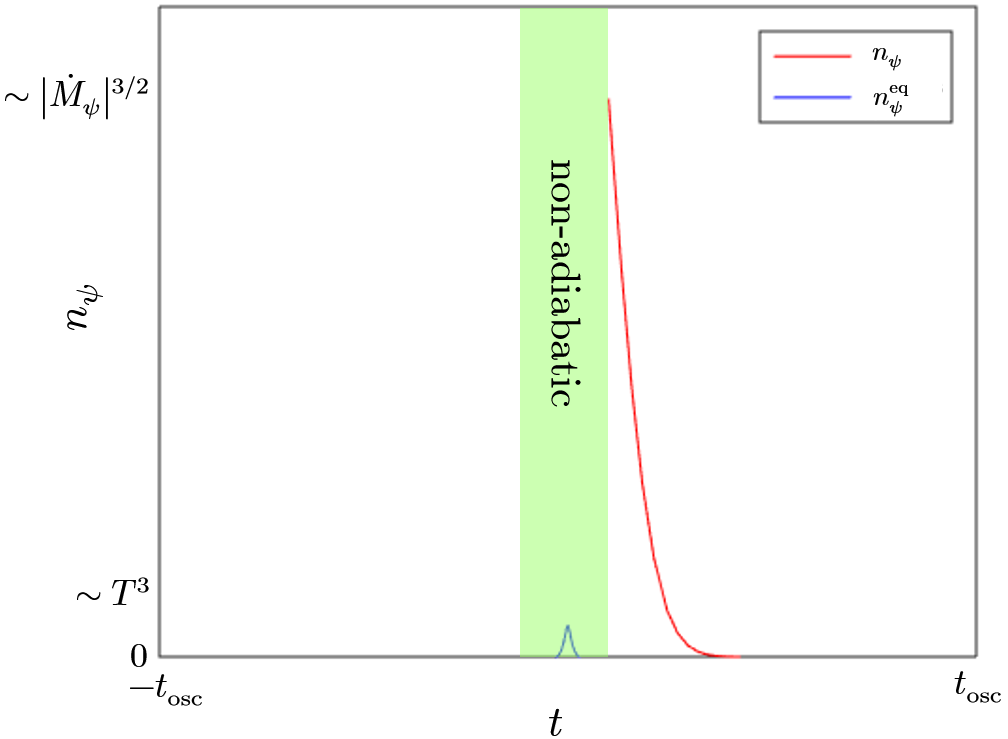}
\caption{Typical evolution of $n_\psi$ during one half-oscillation of $\phi$ in the low-temperature, decay case.}
\label{figure:lowdecay}
\end{figure}

Suppose that $\psi$ starts with zero initial abundance at $\phi_s=-A$. The abundance will remain to be zero until the system encounters the non-adiabatic regime, within which $n^{\text{NA}}_\psi$ of $\psi$ particles per unit volume get created. After the system escapes the non-adiabatic regime, $n_\psi$ evolves according to Eq.~\eqref{ssimboltz}. As $M_\psi$ increases from $|\dot{M}_\psi|^{1/2}$ to $\lambda A$, the decay rate $\Gamma_{\psi\rightarrow\chi\chi}\sim M_\psi$ increases accordingly. A significant fraction of the created $\psi$ particles will decay at around time $t\sim t_{\text{D}}$ that satisfies $\Gamma_{\psi\rightarrow\chi\chi}(t_{\text{D}})t_{\text{D}}\sim 1$. Solving the equation, we get $t_{\text{D}}\sim t_{\text{NA}}<<t_{\text{osc}}$, which means most of the particles will decay when their mass is of order $M_\psi(t_{\text{D}})\sim |\dot{M}_\psi|^{1/2}$. Note that this means parametric resonance does not occur. The energy transferred from $\psi$ to $\chi$ in one half-oscillation is simply the energy of the decaying particle
\begin{align}
\Delta\rho_{\psi\rightarrow\chi}\sim n_{\psi}^{\text{NA}}M_\psi(t_{\text{D}})\sim |\dot{M}_\psi|^2
\end{align}
The next half-oscillation will start with $\psi$ having zero abundance, meaning that the above analysis will still be valid in the subsequent oscillations if the system does not move to a different parameter regime. A typical evolution of $n_\psi$ in this case is shown in Figure~\ref{figure:lowdecay}.

\subsubsection{Annihilation Case}\label{sss1:annihilation case}
\begin{figure}[h]
\centering
\includegraphics[scale=0.45]{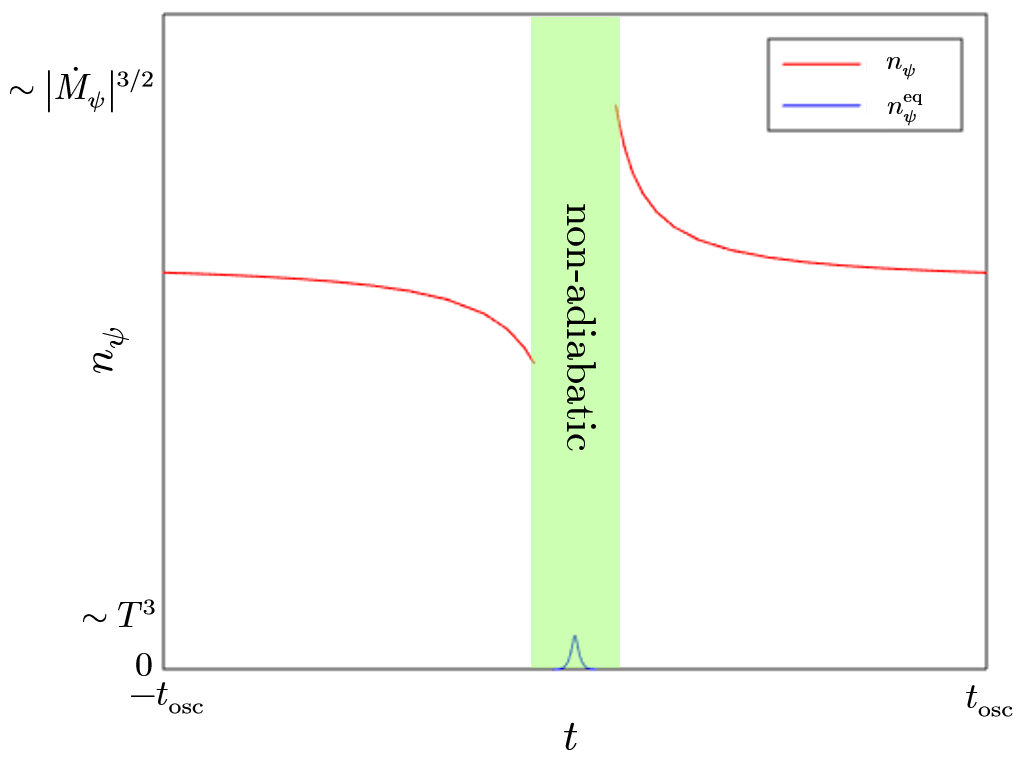}
\caption{Typical evolution of $n_\psi$ during one half-oscillation of $\phi$ in the low-temperature, annihilation case.}
\label{figure:lowannihilation}
\end{figure}

We begin by assuming that $n_\psi=0$ at $\phi_s=-A$. As before, $n_\psi$ will remain zero until non-adiabatic particle production takes place in the non-adiabatic regime. After escaping the non-adiabatic regime, the evolution of $\psi$ is governed by Eq.~\eqref{ssimboltz}. Solving the equation for $t\gtrsim t_{\text{NA}}$, we get
\begin{align}
\frac{1}{n_\psi(t)}&\sim \frac{1}{n_\psi(t_{\text{NA}})}+\frac{1}{|\dot{M}_\psi|M_\psi(t_{\text{NA}})}-\frac{1}{|\dot{M}_\psi|M_\psi(t)}\nonumber\\
&\sim \frac{1}{n_\psi(t_{\text{NA}})}+\frac{1 }{|\dot{M}_\psi|^{3/2}}-\frac{1 }{|\dot{M}_\psi|M_\psi(t)}\label{na2}
\end{align}
Eq.~\eqref{na2} says that after the system escapes the non-adiabatic regime $n_\psi$ will remain to be of order $n_\psi(t_{\text{NA}})\sim |\dot{M}_\psi|^{3/2}$ until the end of the first half-oscillation of $\phi$. The next half-oscillation will start with $n_\psi\sim |\dot{M}_\psi|^{3/2}$ and $n_\psi$ will maintain its order of magnitude until the next non-adiabatic particle production takes place. This time the particle production is Bose-enhanced. So, the new $n_\psi(t_{\text{NA}})$ will be larger than that in the first half-oscillation. However, Eq.~\eqref{na2} shows that no matter how large $n_\psi$ at $t= t_{\text{NA}}$ is, even if it is much larger than $|\dot{M}_\psi|^{3/2}$, it would reduce to around $|\dot{M}_\psi|^{3/2}$ before $M_\psi$ becomes much larger than $|\dot{M}_\psi|^{1/2}$ and remain at that order. Therefore, parametric resonance does not occur, and a generic half-oscillation has the initial condition $n_\psi(t=-t_{\text{osc}})\sim |\dot{M}_\psi|^{3/2}$, as shown in Figure~\ref{figure:lowannihilation}.

Knowing that $n_\psi(t)$ is in a steady state with constant order of magnitude, $n_{\psi}^{\text{NA}}\sim |\dot{M}_\psi|^{3/2}$, we can now calculate the total energy transferred from $\psi$ to $\chi$ in one half-oscillation using Eqs.~\eqref{temperature} and \eqref{ssimboltz}
\begin{align}
\Delta\rho_{\psi\rightarrow\chi}\sim 2\int_{t_{\text{NA}}}^{t_{\text{osc}}} \frac{\left(n_{\psi}^{\text{NA}}\right)^2}{M_\psi^2} M_\psi dt\sim |\dot{M}_\psi|^2\ln\left(\frac{  \lambda A}{|\dot{M}_\psi|^{1/2}}\right)\sim |\dot{M}_\psi|^2\label{annlow}
\end{align}
where we have neglected the contribution from the non-adiabatic regime to $\Delta\rho_{\psi\rightarrow\chi}$, as it is expected to be no bigger than the above result. Even though the particle interpretation does not apply in the non-adiabatic regime, we can estimate the energy transferred from $\psi$ to $\chi$ inside the non-adiabatic regime by integrating Eq.~\eqref{ssimboltz} multiplied by the mass of $\psi$
\begin{align}
\Delta\rho_{\psi\rightarrow\chi}^{\text{NA}}&\sim M_\psi(t_{\text{NA}})\times \frac{\left(n_{\psi}^{\text{NA}}\right)^2}{M_\psi^2(t_{\text{NA}})}\times t_{\text{NA}}\sim |\dot{M}_\psi|^2
\end{align}
which is smaller than or of the same order with the $\Delta\rho_{\psi\rightarrow\chi}$ found in Eq.~\eqref{annlow}.

\subsection{Intermediate Temperature Regime: \boldmath $|\dot{M}_\psi|^{1/2}<<T<<\lambda A$}\label{ss:intermediate temperature regime}
In this case, non-adiabatic particle production is negligible and the non-adiabatic regime is much smaller than the thermal regime. Also, the condition $T<<\lambda A$ implies that the thermal regime is small compared to the span of the possible values of $M_\psi$. We will see that this allows the freeze out phenomenon to occur in the annihilation case. Furthermore, the condition $T>>|\dot{M}_\psi|^{1/2}$ means that $\psi$ can closely track its equilibrium abundance inside the thermal regime. Subtracting $\dot{n}_\psi^{\text{eq}}$ on both sides of Eq.~\eqref{simboltz}, we get
\begin{align}
\delta \dot{n}_\psi\sim-\Gamma_{\text{th}}\delta n_\psi-\dot{n}_\psi^{\text{eq}} \label{approxboltzmann}
\end{align}
where we have defined $\delta n_\psi=n_\psi-n_\psi^{\text{eq}}$. When $n_\psi$ is tracking $n_\psi^{\text{eq}}$ closely, $\delta \dot{n}_\psi(t)$ is small compared to $\dot{n}_\psi^{\text{eq}}$. Setting the left hand side of Eq.~\eqref{approxboltzmann} to zero, we obtain
\begin{align}
\delta n_\psi\sim -\frac{1}{\Gamma_{\text{th}}}\dot{n}_\psi^{\text{eq}}\label{stick}
\end{align}
where $\dot{n}_\psi^{\text{eq}}$ can be calculated as follows
\begin{align}
\dot{n}_\psi^{\text{eq}}&=\int \frac{d^3k}{(2\pi)^3}\frac{d}{dt}\left(\frac{1}{e^{\sqrt{M_\psi^2+k^2}/T}+1}\right)\nonumber\\
&\sim -\frac{M_\psi \dot{M}_\psi}{T}\int d^3k \frac{e^{\sqrt{M_\psi^2+k^2}/T}}{\sqrt{M_\psi^2+k^2}\left(e^{\sqrt{M_\psi^2+k^2}/T}+1\right)^2}\nonumber\\
&\sim \begin{cases}
\displaystyle   \hfil-M_\psi T \dot{M}_\psi       & \text{ for }  M_\psi<< T\\[2ex]
\displaystyle   \hfil-M_\psi^{3/2}T^{1/2}\dot{M}_\psi e^{-M_\psi/T}  & \text{ for } M_\psi >> T
  \end{cases}
\end{align}
Eq.~\eqref{stick} says that if $n_\psi^{\text{eq}}$ is increasing then $n_\psi$ will lag below it and if $n_\psi^{\text{eq}}$ is decreasing then $n_\psi$ will lag above it, as expected.

The qualitative behaviour of $n_\psi$ after exiting the thermal regime is different in the decay and annihilation case. In the decay case, $n_\psi$ can continue to track its equilibrium value without any problem as it decreases to zero because its decay rate is independent of $n_\psi$ and increases with $M_\psi$. In the annihilation case, each of the $\psi$ particles has to find another $\psi$ particle in order to annihilate. So, as $n_\psi$ decreases, at some point the annihilation rate of $\psi$ becomes so slow that $n_\psi$ is no longer able to track $n_\psi^{\text{eq}}$. We will refer to this point as the \textit{thermal decoupling} point. Then, $n_\psi$ will continue to decrease until at some point the annihilation rate of $\psi$ becomes so small that $n_\psi$ essentially stops decreasing. When that happens we say that $\psi$ \textit{freezes out} \citep{Kolb:1990vq}.

\subsubsection{Decay Case}\label{sss2:decay case}
\begin{figure}[h]
\centering
\includegraphics[scale=0.45]{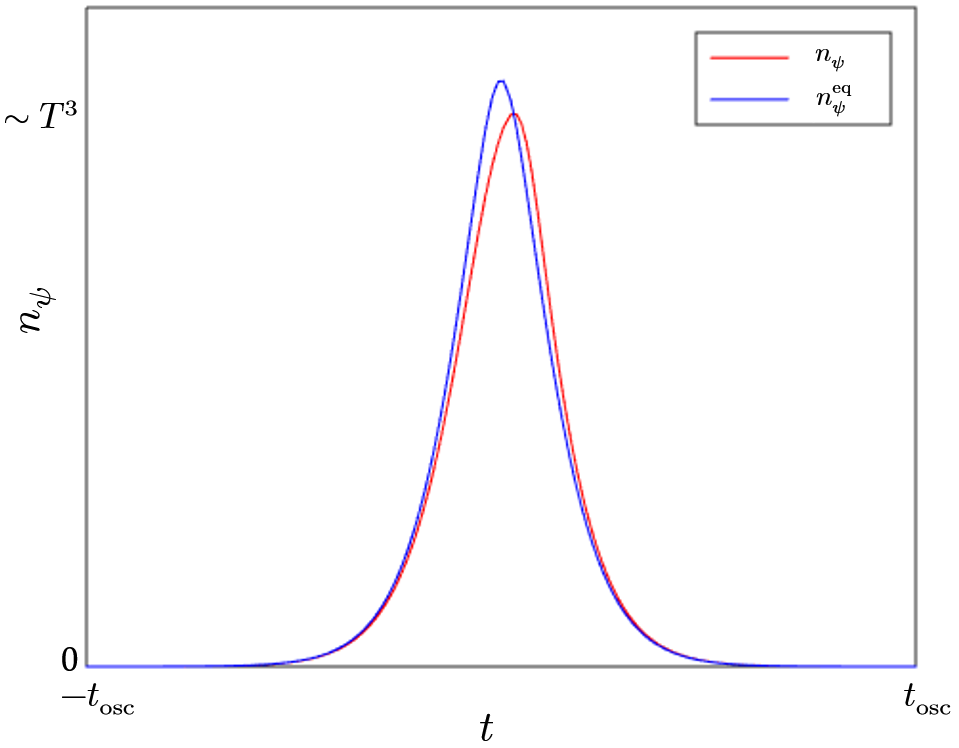}
\caption{Typical evolution of $n_\psi$ during one half-oscillation of $\phi$ in the intermediate-temperature, decay case.}
\label{figure:intermediatedecay}
\end{figure}

Let us start by assuming that the initial abundance of $\psi$ at $\phi=-A$ is zero. The abundance will remain close to zero until the system enters the thermal regime, where $n_\psi$ will closely follow the unsuppressed $n_\psi^{\text{eq}}$ while lagging behind by $\delta n_\psi$, given by Eq.~\eqref{stick}. After escaping the thermal regime, $n_\psi$ will still closely follow $n_\psi^{\text{eq}}$ and get Boltzmann suppressed. Therefore, $n_\psi$ will decay completely way before the end of the half-oscillation, justifying the assumption that the initial abundance of $\psi$ can be taken to be zero. Typical evolution of $n_\psi$ in this case is shown in Figure~\ref{figure:intermediatedecay}.

Since we have $n_\psi\approx0$ at the start and end of the half-oscillation, the energy input and output of $\psi$ must balance. The energy density transferred from $\psi$ to $\chi$ in one half-oscillation of $\phi$ can thus be calculated using Eq.~\eqref{damp}
\begin{align}
\Delta \rho_{\psi\rightarrow\chi}=\Delta \rho_{\phi\rightarrow\psi}&\sim\int_{-t_{\text{osc}}}^{t_{\text{osc}}} \left(n_\psi^{\text{eq}}+\delta n_\psi\right) \dot{M}_\psi dt\sim |\dot{M}_\psi|T^2 \label{thregimeonly}
\end{align}
In the last step, we neglected the term involving $n_\psi^{\text{eq}}$ because $n_\psi^{\text{eq}}$ is even while $\dot{M}_\psi$ is odd. Furthermore, the integration result in equation \eqref{thregimeonly} comes mainly from the thermal regime since $\delta n_\psi$ is Boltzmann suppressed outside the thermal regime. Since $|\dot{M}_\psi|^{1/2}<<T$, $\Delta\rho_{\psi\rightarrow\chi}$ found in Eq.~\eqref{thregimeonly} is clearly much smaller than $\rho_\chi\sim g_* T^4$, meaning that our assumption of neglecting the temperature variation during one half-oscillation is justified.

\subsubsection{Annihilation Case}\label{sss2:annihilation case}
As before, assuming that $n_\psi$ is initially zero at $\phi=-A$, it will remain so until the system encounters the thermal regime, within which $\psi$ can be thermally excited. As the system is escaping the thermal regime, $n_\psi$ will continue tracking $n_\psi^{\text{eq}}$ for a while, but, at some point, say at $t\sim t_{\text{TD}}$, $n_\psi$ starts to deviate significantly from $n_\psi^{\text{eq}}$. To be concrete, we define the time of thermal decoupling $t_{\text{TD}}$ as the time when 
\begin{align}
\delta n_\psi(t_\text{TD})= \xi n_\psi^{\text{eq}}(t_\text{TD}) \label{separation}
\end{align}
Where $\xi$ is an $O(1)$ constant whose exact value is not important. It is difficult to determine $t_{\text{TD}}$ precisely, but the results of our calculations are insensitive to the exact value of $t_{\text{TD}}$. We, at least, know that 
\begin{align}
t_{\text{TD}}&\sim t_{\text{th}}=\frac{T}{|\dot{M}_\psi|}\\
M_\psi(t_{\text{TD}})&\sim T \label{separationmass}
\end{align}
because there is where the Boltzmann suppression starts to kick in. It can be shown (Appendix \ref{s:point of separation}) that at the time of thermal decoupling
\begin{align}
n_\psi(t_{\text{TD}}) \sim |\dot{M}_\psi|T \label{sdensity} 
\end{align}

\begin{figure}[h]
\centering
\includegraphics[scale=0.45]{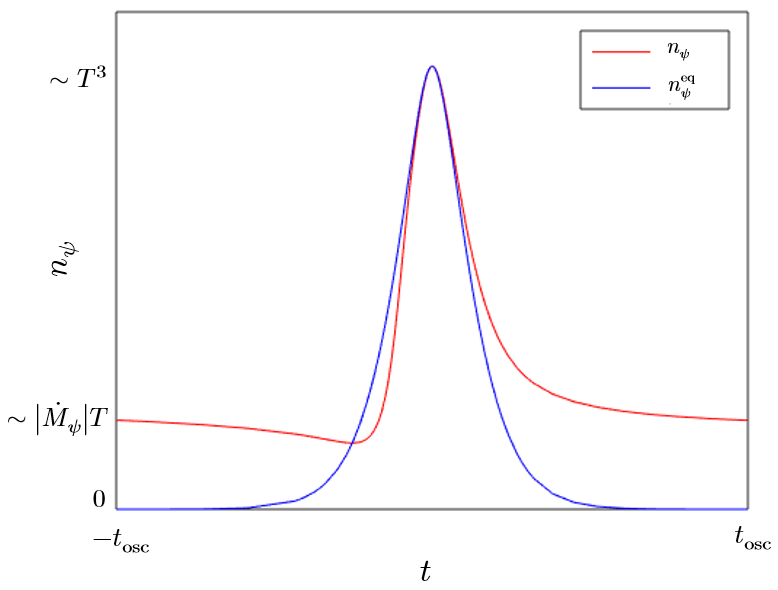}
\caption{Typical evolution of $n_\psi$ during one half-oscillation of $\phi$ in the intermediate-temperature, annihilation case.}
\label{figure:intermediateannihilation}
\end{figure}

After $t_{\text{TD}}$, the system encounters a regime where $n_{\psi}$ has departed from $n_\psi^{\text{eq}}$ but $\delta n_\psi$ is still of order $n_\psi^{\text{eq}}$. The evolution of $n_\psi$ in this regime is complicated, so we make the following approximation. We calculate the evolution of $n_\psi$ after $t_{\text{TD}}$ using Eq.~\eqref{ssimboltz}, i.e.\ we set $n_\psi^{\text{eq}}=0$. Neglecting $n_\psi^{\text{eq}}$ here will not change the order of magnitude of our results, because when we do so we only overestimate the annihilation rate of $\psi$ by some $O(1)$ factor. The solution of Eq.~\eqref{ssimboltz} for $t\geq t_{\text{TD}}$ is
\begin{align}
\frac{1}{n_\psi(t)}&\sim \frac{1}{n_\psi(t_{\text{TD}})}+\frac{1 }{|\dot{M}_\psi|}\left[\frac{1}{M_\psi(t_{\text{TD}})}-\frac{1}{M_\psi(t)}\right] \label{ssimboltzsol}
\end{align}
Therefore using Eq.~\eqref{sdensity}
\begin{align}
n_\psi(t)&\sim |\dot{M}_\psi|T\left/ \left\{1+T\left[\frac{1}{M_\psi(t_{\text{TD}})}-\frac{1}{M_\psi(t)}\right]\right\}\right.
\label{sol}
\end{align}
Using Eq.~\eqref{separationmass}, the solution tells us that if $\lambda A>>T$ the abundance of $\psi$ at late times will freeze out at the value 
\begin{align}
n_\psi^{\text{F}}\sim |\dot{M}_\psi| T  \label{fdensity} 
\end{align}
Thus, a generic half-oscillation of is one where $\psi$ starts with a freeze out density $n_\psi^{\text{F}}$, a leftover from the previous half-oscillation, as shown in Figure~\ref{figure:intermediateannihilation}.

\begin{figure}[h]
\centering
\includegraphics[scale=0.425]{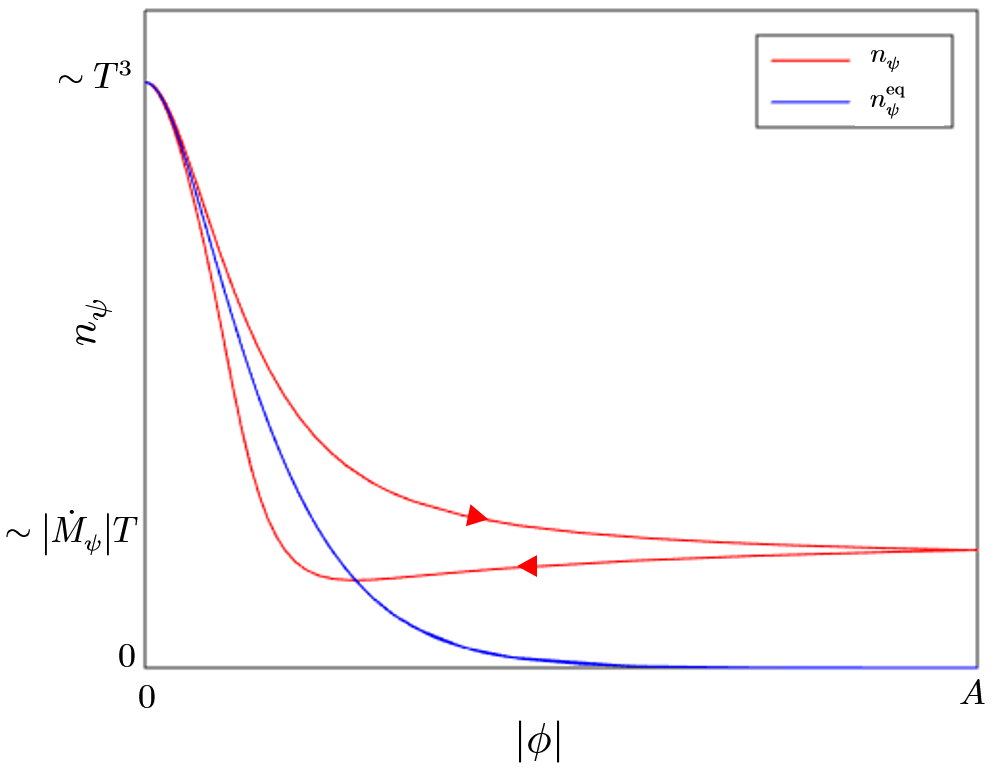}
\caption{Figure~\ref{figure:intermediateannihilation} plotted with respect to $|\phi|$. The arrows indicate the direction of evolution.}
\label{figure:intermediateannihilation2}
\end{figure}

To calculate the amount of mediated damping, it is more convenient to consider a half-oscillation of $\phi$ that starts from $\phi=0$, goes to $|\phi|=A$, and ends at $\phi=0$. In Figure~\ref{figure:intermediateannihilation2}, we re-plot Figure~\ref{figure:intermediateannihilation} with respect to $|\phi|$ in order to make it a better aid for our calculation and to emphasize the difference in $n_\psi$ during the ascending and descending parts of $\phi$'s oscillation. As we can see in Figure~\ref{figure:intermediateannihilation2}, at $|\phi|=A$ we have $n_\psi^{\text{asc}}=n_\psi^{\text{des}}$, and as we decrease $|\phi|$, $n_\psi^{\text{asc}}$ and $n_\psi^{\text{des}}$ will depart from one another. While $n_\psi^{\text{asc}}-n_\psi^{\text{des}}<<n_\psi^{\text{F}}$, integrating Eq.~\eqref{ssimboltz} gives
\begin{align}
n_\psi^{\text{asc}}-n_\psi^{\text{des}}&\sim -2\times \int_{\lambda A}^{M_\psi} \frac{\left(n_\psi^{\text{F}}\right)^2}{M_\psi^2} \frac{dM_\psi}{|\dot{M}_\psi|}\sim |\dot{M}_\psi|T^2\left(\frac{1}{M_\psi}-\frac{1}{\lambda A}\right)\label{approx1}
\end{align}
which is valid as long as $M_\psi\gtrsim T$, and so is applicable for $t\geq t_{\text{TD}}$.

The energy transferred from $\psi$ to $\chi$ via mediated damping in one half-oscillation of $\phi$ can be conveniently calculated by rewriting Eq.~\eqref{damp} in terms of an integral over the effective mass $M_\psi$
\begin{align}
\Delta\rho_{\psi\rightarrow\chi}\approx\Delta\rho_{\phi\rightarrow\psi}&\sim\int_{m_{\psi,\text{th}}}^{\lambda A} \left(n_\psi^{\text{asc}}-n_\psi^{\text{des}}\right) dM_\psi \nonumber\\ 
&=\int_{m_{\psi,\text{th}}}^{M_\psi(t_{\text{TD}})} \left(n_\psi^{\text{asc}}-n_\psi^{\text{des}}\right)dM_\psi+\int_{M_\psi(t_{\text{TD}})}^{\lambda A} \left(n_\psi^{\text{asc}}-n_\psi^{\text{des}}\right)dM_\psi \nonumber\\ 
&\sim 2\int_{m_{\psi,\text{th}}}^{M_\psi(t_{\text{TD}})} |\delta n_\psi| dM_\psi+2\int_{M_\psi(t_{\text{TD}})}^{\lambda A} |\dot{M}_\psi|T^2\left(\frac{1}{M_\psi}-\frac{1}{\lambda A}\right) dM_\psi\nonumber\\
&\sim |\dot{M}_\psi|T^2+|\dot{M}_\psi|T^2\left[\ln\left(\frac{\lambda A}{T}\right)-1\right]\nonumber\\
&\sim |\dot{M}_\psi|T^2 \label{dampedintermediateannihilation}
\end{align}
where Eqs.~\eqref{approx1} and \eqref{stick} are employed in going from the second to third line and in going from the third to fourth line respectively. The obtained expression for $\Delta\rho_{\psi\rightarrow\chi}$ is the same as in the decay case, so we again find that the change in the temperature of the thermal bath 
\begin{align}
\Delta T\sim \frac{|\dot{M}_\psi|}{g_* T}
\end{align}
is small in the sense that $\Delta T<<T$. As a consequence of this relatively small change in $T$, the freeze out density of $\psi$ in the next half-oscillation will increase by $\Delta n_\psi^{\text{F}}\sim |\dot{M}_\psi| \Delta T$. This means that about
\begin{align}
\lambda A\Delta n_\psi^{\text{F}}\sim \frac{\lambda A |\dot{M}_\psi|^2}{g_* T} \label{increaseintrapping}
\end{align}
more energy per unit volume will end up in the $\psi$ sector when $|\phi|=A$, which, as discussed in Section~\ref{s:mediated damping process}, contributes to decreasing $A$ alongside other damping effects. If Eq.~\eqref{increaseintrapping} is larger than the $\Delta\rho_{\psi\rightarrow\chi}$ found in Eq.~\eqref{dampedintermediateannihilation}, then $A$ will decrease more because of trapping than damping.

\subsection{High Temperature Regime: \boldmath $\lambda A<<T$}\label{ss:high temperature regime}
\begin{figure}[h]
\centering
\includegraphics[scale=0.45]{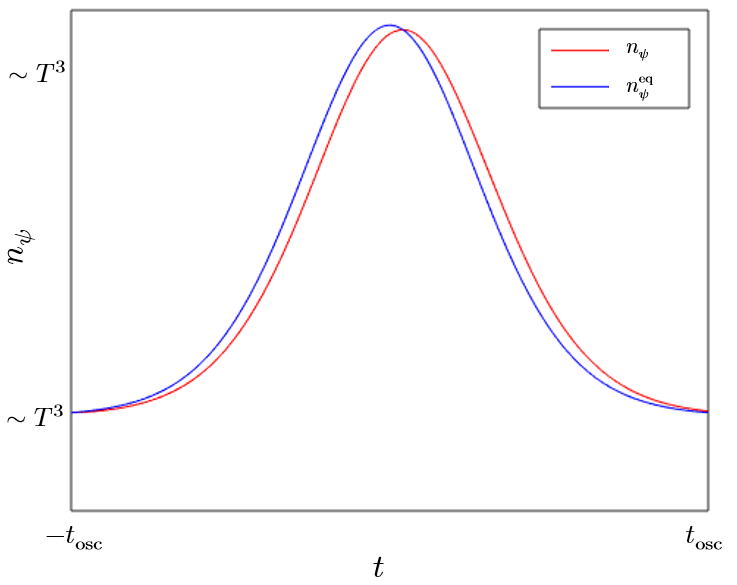}
\caption{Typical evolution of $n_\psi$ during one half-oscillation of $\phi$ in the high-temperature, decay/annihilation case.}
\label{figure:high}
\end{figure}

In this case, $\psi$ never decouples from the thermal bath and $M_\psi$ is always of order $T$. Following the same argument as in the intermediate temperature regime, the deviation of $n_\psi$ from its thermal equilibrium value is given by Eq.~\eqref{stick}
\begin{align}
\delta n_\psi\sim -\frac{\dot{n}_\psi^{\text{eq}}}{\Gamma_{\text{th}}}\sim \dot{M}_\psi T \label{veryhighdeviation}
\end{align}
which applies both in the decay and annihilation case. The energy density transferred to $\chi$ after one half-oscillation can be calculated using Eq.~\eqref{damp} as
\begin{align}
\Delta\rho_{\psi\rightarrow\chi}\approx \Delta\rho_{\phi\rightarrow\psi}&\sim  \int_{-t_{\text{osc}}}^{t_{\text{osc}}}\delta n_\psi \dot{M}_\psi dt\sim |\dot{M}_\psi|\left(\lambda A\right)^2 \label{veryhighdamp}
\end{align}
Typical evolution of $n_\psi$ in this case is shown in Figure~\ref{figure:high}.

\section{Effective Potential of \boldmath $\phi$}\label{s:effective potential of phi}
The rate the energy of $\phi$ is damped is given by the energy transferred to $\chi$ per $\phi$-oscillation, $\Delta \rho_{\psi\rightarrow\chi}$, calculated in the previous section, divided by the oscillation period $t_{\text{osc}}$. In turn, $t_{\text{osc}}$ depends on the effective potential $V_{\text{eff}}(\phi)$ of $\phi$, which can be significantly modified if $\psi$ is sufficiently excited. Both $\Delta \rho_{\psi\rightarrow\chi}$ and $V_{\text{eff}}(\phi)$, and thus $t_{\text{osc}}$, are determined by the dynamics of $\psi$, which is driven by the time-dependent mass it receives from $\phi$ and its tendency to be thermalized by $\chi$. To characterize such dynamics, it is sufficient to specify three scales: the rate at which the mass of $\psi$ changes $|\dot{M}_\psi|^{1/2}$, the temperature of the thermal bath $T$, and $\lambda A$, which determines the amplitude of the $\psi$ mass variation. One of the three scales, $|\dot{M}_\psi|^{1/2}$, is not a free parameter. It is determined in a non-trivial way once $T$ and $\lambda A$ are given. Therefore, to evaluate how the amplitude $A$ of $\phi$ and the temperature $T$ of $\chi$ change with time, it is sufficient to keep track of $T$ and $\lambda A$. These two quantities will constitute the parameter space of our system. One of our assumptions, Eq.~\eqref{amplitude}, limits our exploration to the $\lambda A>>m_\phi$ region of this parameters space.

\begin{figure}
\centering
\includegraphics[scale=0.35]{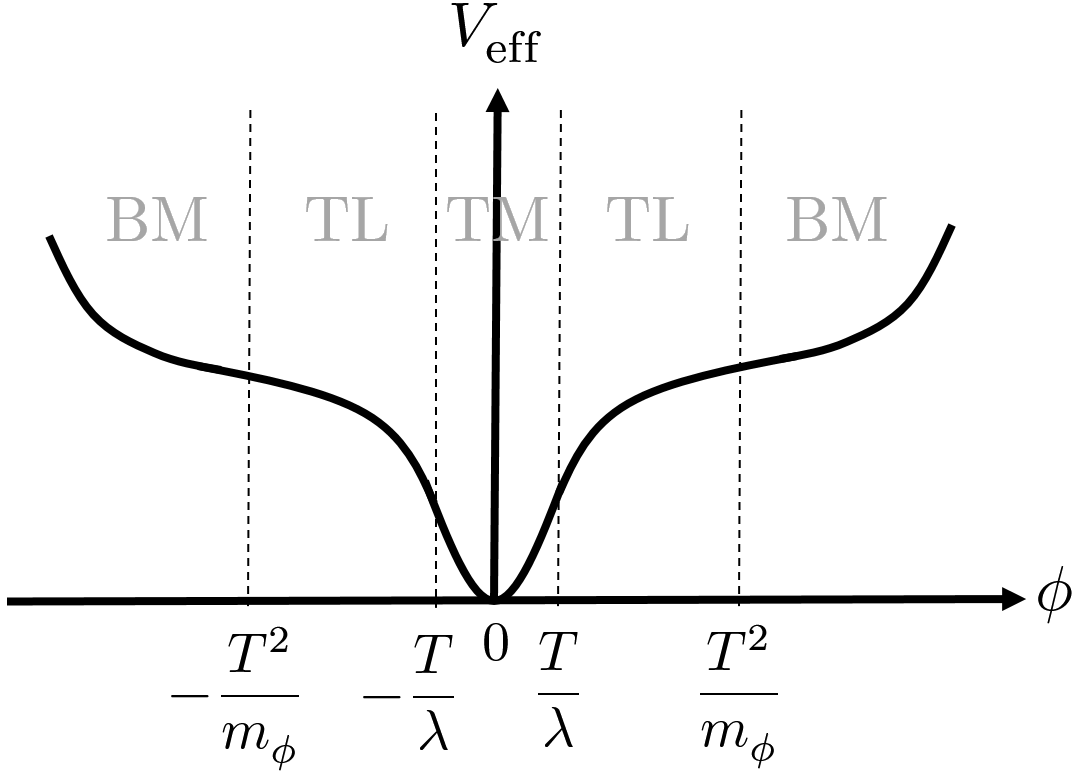}
\caption{Full form of $V_{\text{eff}}(\phi)$ in the decay case with $\lambda T>>m_\phi$. In the annihilation case, there may be an additional linear potential arising from the frozen out $\psi$ particles' back-reaction on $\phi$.}
\label{figure:example}
\end{figure}

 In this section, we will map out the parameter space by determining the form of $V_{\text{eff}}(\phi)$ throughout the parameter space, and the $|\dot{M}_\psi|$ it gives for given $T$ and $\lambda A$. We have seen in the previous section that the damping process is different depending on the size of $T$ relative to $|\dot{M}_\psi|^{1/2}$ and $\lambda A$. For each $\lambda A$, the parameter space is divided into three regions: low temperature region with $|\dot{M}_\psi|^{1/2}>>T$, intermediate temperature region with $|\dot{M}_\psi|^{1/2}<<T<<\lambda A$, and high temperature region with $T>>\lambda A$, based on the nature of the $\psi$ dynamics. In this section, we find that the parameter space can be further divided based on the character of $V_{\text{eff}}(\phi)$. As shown in Figure~\ref{figure:summary}, the parameter space can be divided into: \textit{bare mass}, \textit{thermal mass}, \textit{thermal log}, and \textit{linear potential} (annihilation case) regions. The actual form of $V_{\text{eff}}(\phi)$ is some combination of these potentials, although typically one form dominates over the others. For example, Figure~\ref{figure:example} shows $V_{\text{eff}}(\phi)$ in the decay case when $\lambda T>>m_\phi$.

\subsection{Bare Mass }\label{ss:bare mass}
If the back-reaction from $\psi$ to $\phi$ is negligible
\begin{align}
\lambda^2\left<\psi^2\right><<m_\phi^2
\end{align}
$\phi$ oscillates with its bare mass 
\begin{align}
V_{\text{eff}}^{\text{BM}}(\phi)=\frac{1}{2}m_\phi^2\phi^2
\end{align}
and
\begin{align}
|\dot{M}_\psi|=\lambda|\dot{\phi}|\frac{\lambda \phi}{M_\psi}\sim \lambda Am_\phi\frac{\lambda A}{\max\left(\lambda A,m_{\psi,\text{th}}\right)}\sim \lambda Am_\phi \min\left(1,\frac{\lambda A}{T}\right)
\end{align}

\subsection{Thermal Mass}\label{ss:thermal mass}
When $\psi$ is in a thermal state without Boltzmann suppression at temperature $T$, it induces a thermal mass potential on $\phi$
\begin{align}
V_{\text{eff}}^{\text{TM}}(\phi)\sim\lambda^2T^2\phi^2
\end{align}
In the high temperature regime $T>>\lambda A$, where $\psi$ remains in a thermal state without Boltzmann suppression, $\phi$ is trapped by the thermal mass potential if $\lambda T\gtrsim m_\phi$, which can be rewritten as
\begin{align}
\frac{T}{m_\phi}\gtrsim\frac{1}{\lambda}
\end{align}
If the above condition is satisfied, $|\dot{M}_\psi|$ is given by
\begin{align}
|\dot{M}_\psi|=\lambda|\dot{\phi}|\frac{\lambda \phi}{M_\psi}\sim \lambda^3A^2
\end{align}

\subsection{Thermal Log}\label{ss:thermal log}
When $T<<\lambda A$ and the $\psi$ particles are scarce, although the direct effect of the scarce $\psi$ particles is small, the potential of $\phi$ still receives corrections due to its $\psi$-mediated, higher loop order interactions with the thermal bath $\chi$. This effect results in an extra \textit{thermal log potential} \citep{Anisimov2001,Mukaida2012,Mukaida2013}
\begin{align}
V_{\text{eff}}^{\text{TL}}(\phi)\sim  T^4\ln\left(\frac{\lambda\phi}{T}\right)
\end{align}
which is valid when $\lambda \phi>>T$ and might alter the motion of the $\phi$ field. The thermal log potential can store $\sim T^4$ of energy. Thus, the thermal log potential dominates over the bare mass potential if
\begin{align}
T^4\gtrsim \left.\frac{1}{2}m_\phi^2\phi^2\right|_{|\phi|=A}
\end{align}
or
\begin{align}
T^2&\gtrsim m_\phi A
\end{align}
If the thermal log potential dominates, then $\dot{\phi}^2\sim T^4$, and $|\dot{M}_\psi|$ is modified to
\begin{align}
|\dot{M}_\psi|\sim \lambda\dot{\phi}\sim \lambda  T^2 
\end{align}

\subsection{Linear Potential (Annihilation Case)}\label{ss:linear potential}
In the annihilation case with $T<<\lambda A$ (low and intermediate temperature regime), the $\psi$ particles never completely disappear because their annihilation rate slows down as their number density decreases. The leftover particles may store a significant amount of energy density when their effective mass gets big, effectively giving a \textit{linear potential} to $\phi$ \citep{beauty, Moroi2013}
\begin{align}
\fn{V_{\text{eff}}^{\text{LP}}}{\phi}=  n_\psi^{\text{F}}M_\psi\sim \lambda|\dot{M}_\psi| \max\left(|\dot{M}_\psi|^{1/2},T\right)|\phi| \label{linearpotential}
\end{align}
where the freeze out abundance $n_\psi^{\text{F}}$ is given by Eq.~\eqref{NAparticle} in the low temperature region $T<<|\dot{M}_\psi|^{1/2}$ and Eq.~\eqref{fdensity} in the intermediate temperature region $|\dot{M}_\psi|^{1/2}<<T<<\lambda A$. $V_{\text{eff}}^{\text{LP}}(\phi)$, in turn, will determine the way $\phi$ oscillates and thus $|\dot{M}_\psi|$. Solving $|\dot{M}_\psi|\sim \lambda |\dot{\phi}|$ and $\dot{\phi}^2\sim \left.\fn{V_{\text{eff}}^{\text{LP}}}{\phi}\right|_{|\phi|=A}$ for $|\dot{M}_\psi|$ gives

\begin{align}
|\dot{M}_\psi|\sim\begin{cases}
\displaystyle \hfil \lambda^6A^2 &\text{for }T<<|\dot{M}_\psi|^{1/2} \phantom{abc}\text{i.e. }T<<\lambda^3 A\\[2ex]
\displaystyle \hfil \lambda^3 AT &\text{for }T>>|\dot{M}_\psi|^{1/2} \phantom{abc}\text{i.e. }T>>\lambda^3 A
\end{cases}\label{linpotmpsidot}
\end{align}

When the $\psi$ number density $n_\psi$ freezes at a certain value, $\psi$ is pretty much decoupled from the thermal bath. So, when $V_{\text{eff}}^{\text{LP}}(\phi)$ increases, energy must be drained from $\phi$. If $V_{\text{eff}}^{\text{LP}}(\phi)$ drains too much energy from $\phi$, it may significantly reduce the amplitude of $\phi$, i.e.\ $\phi$ is linearly trapped. The amplitude of $\phi$ is significantly modified if the linear potential is dominant or comparable to the bare mass potential of $\phi$ at its oscillation amplitude $|\phi|= A$
\begin{align}
\left.V_{\text{eff}}^{\text{LP}}(\phi)\right|_{|\phi|=A}\gtrsim\left.\frac{1}{2}m_\phi^2\phi^2\right|_{|\phi|=A} \label{lintrap}
\end{align}
Using Eq.~\eqref{linearpotential}, Eq~\eqref{linpotmpsidot}, and Eq.~\eqref{lintrap} we get the linear potential-bare mass boundary
\begin{align}
\frac{\lambda A}{m_\phi}&\gtrsim \frac{1}{\lambda^4} \phantom{abc}\text{for } T<<\lambda^3A\nonumber\\
\frac{T}{m_\phi}&\gtrsim \frac{1}{\lambda^2} \phantom{abc}\text{for } T>>\lambda^3A
\end{align}
Furthermore, the linear potential-thermal log boundary is given by
\begin{align}
T^4&\gtrsim \left.V_{\text{eff}}^{\text{LP}}(\phi)\right|_{|\phi|=A}
\end{align}
where $V_{\text{eff}}^{\text{LP}}(\phi)$ is given by Eq.~\eqref{linearpotential}, i.e.
\begin{align}
\frac{T}{\lambda A}&\gtrsim \lambda
\end{align}

\begin{figure}
\centering
\includegraphics[scale=0.45]{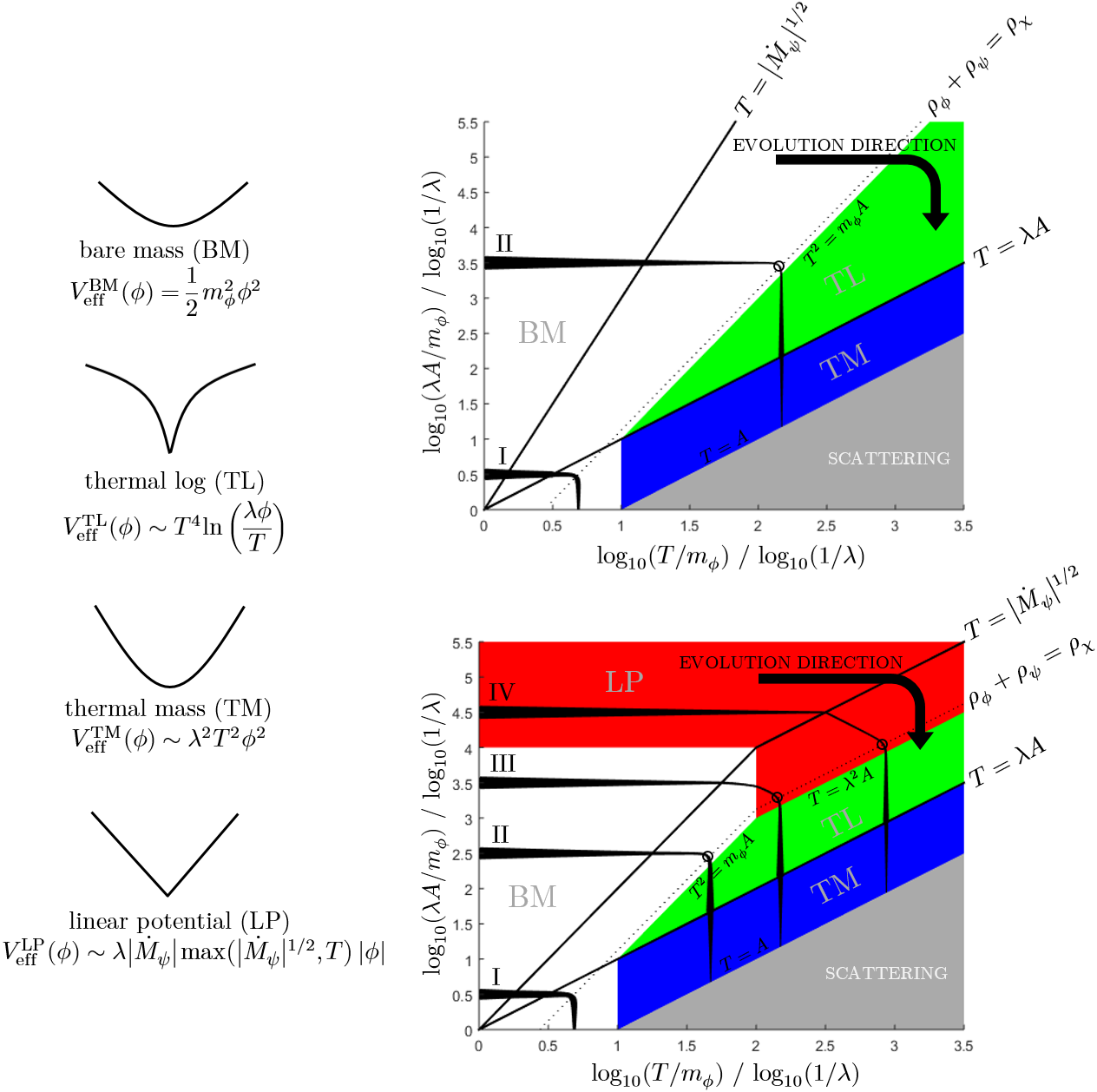}
\caption{Parameter space diagram of the system in the decay (top) and annihilation (bottom) case.
\textbf{Colored regions} indicate types of $\phi$ effective potential: \fcolorbox{black}{white}{\rule{0pt}{6pt}\rule{6pt}{0pt}} bare mass (BM), \fcolorbox{black}{blue}{\rule{0pt}{6pt}\rule{6pt}{0pt}} thermal mass (TM), \fcolorbox{black}{green}{\rule{0pt}{6pt}\rule{6pt}{0pt}} thermal log (TL), and \fcolorbox{black}{red}{\rule{0pt}{6pt}\rule{6pt}{0pt}} linear potential (LP). \textbf{Bold solid lines} divide these regions further based on the nature of $\psi$ dynamics into low temperature $T<<|\dot{M}_\psi|^{1/2}$, intermediate temperature $|\dot{M}_\psi|^{1/2}<<T<<\lambda A$, and high temperature $\lambda A<<T$ regions. \textbf{Lines with varying thickness} show possible trajectories of the system with their thickness representing instantaneous evolution rate ($\chi$ heating rate or $\phi$ damping rate); each representative trajectory is marked with a roman numeral. \textbf{Small circles} indicate local minima of the evolution rate, the choke points of the process. \textbf{Bent arrows at the corners} show the direction of evolution along a trajectory, i.e.\ the system generally evolves towards increasing $T$ and decreasing $A$ direction. After a trajectory crosses the $\rho_\phi+\rho_\psi=\rho_\chi$ \textbf{dotted line}, $T$ stops growing significantly and the system moves predominantly in the decreasing $A$ direction. The colored regions apply for all $\lambda$'s, but the depicted trajectories correspond to $\lambda=10^{-4}$. In the \fcolorbox{black}{gray}{\rule{0pt}{6pt}\rule{6pt}{0pt}} regions, where scattering dominates the damping process and $\phi$ fluctuations grow, the diagram no longer describes the system adequately.}
\label{figure:summary}
\end{figure}

\section{Results}\label{s:results}
\subsection{Parameter Space Trajectory (Kinematics)}\label{ss:kinematics}
Regardless of the details of $\phi$ damping, we can deduce the following about the trajectory of the system in the parameter space diagram (Figure~\ref{figure:summary}).

First, we consider the decay case. Let us follow a trajectory of the system in the parameter space starting from the point where the $\phi$ field has a large amplitude $A_\mathrm{i}$ and the thermal bath $\chi$ has an extremely low temperature. In the beginning, $\phi$ has much more energy than $\chi$ does, and so it would take a lot more energy transfer to decrease the $\phi$ amplitude $A$ than to increase the thermal bath temperature $T$. Thus, the system will initially move in the increasing $T$ direction in the parameter space. It will continue moving in this manner until the energy density of $\phi$ and $\chi$ become comparable. At around this point, $A$ begins to decrease significantly and $T$ begins to asymptotically approach its maximum value
\begin{align}
T_{\max}^{\text{D}}\sim \left(\frac{\rho_{\phi,\text{i}}}{ g_* }\right)^{1/4}\sim\frac{A_\mathrm{i}^{1/2} m_\phi^{1/2}}{ g_*^{1/4} } \label{Tmaxdecay}
\end{align}
The trajectory then has no choice but to bend towards the decreasing $A$ direction and continue moving that way as $T$ can no longer increase significantly.

In the annihilation case, there is an extra complication that the amplitude of $\phi$ may decrease more because of increasing trapping effect than because of damping effect. This only happens in the thermal linear potential region, in which the trapping effect gets stronger as $T$ increases. As a consequence, the parameter space trajectory may start to curve partially towards the decreasing $A$ direction before the temperature is close to its maximum value. The trajectory will finally fully bend towards the decreasing $A$ direction when $T\simeq T_{\max}$. The $T_{\max}$ corresponding to a particular $A_\mathrm{i}$ may differ in the decay and annihilation case. The reason is that if the system starts in the linear trapping region in the annihilation case, the total energy of the system would initially be dominated by $\psi$ instead of $\phi$. Hence, the maximum temperature of $\chi$ in the annihilation case is given by
\begin{align}
T_{\max}^{\text{A}}&\sim \max\left[ \left(\frac{\rho_{\phi,\text{i}}}{ g_* }\right)^{1/4}, \left(\frac{\rho_{\psi,\text{i}}}{ g_* }\right)^{1/4}\right]\sim \max\left[\frac{A_\mathrm{i}^{1/2} m_\phi^{1/2}}{ g_*^{1/4} }, \frac{\lambda^{5/2}A_\mathrm{i}}{g_*^{1/4}}\right] \label{Tmaxannihilation}
\end{align}

Due to the conservation of energy, there is a unique trajectory in the parameter space corresponding to each possible value of the total energy of the system. As shown in Figure~\ref{figure:summary}, we can identify four different ways in which the trajectory of the system cuts through the different damping regimes. They can be classified by their initial amplitudes $A_\mathrm{i}$ as
\begin{align}
\setlength\arraycolsep{1pt}
\begin{array}{LRL}
\text{Type I:  } &&\frac{\lambda A_\mathrm{i}}{m_\phi}<<\frac{1}{g_*^{1/2}\lambda}\\[2ex]
\text{Type II:  } &\frac{1}{g_*^{1/2}\lambda}<<&\frac{\lambda A_\mathrm{i}}{m_\phi}
\end{array}\label{typesd}
\end{align}
in the decay case and
\begin{align}
\setlength\arraycolsep{1pt}
\begin{array}{LRL}
\text{Type I:  } &&\frac{\lambda A_\mathrm{i}}{m_\phi}<<\frac{1}{g_*^{1/2}\lambda}\\[2ex]
\text{Type II:  } &\frac{1}{g_*^{1/2}\lambda}<<&\frac{\lambda A_\mathrm{i}}{m_\phi}<<\frac{g_*^{1/2}}{\lambda^3}\\[2ex] 
\text{Type III:  } &\frac{g_*^{1/2}}{\lambda^3}<<&\frac{\lambda A_\mathrm{i}}{m_\phi}<<\frac{1}{\lambda^4}\\[2ex]
\text{Type IV:  } &\frac{1}{\lambda^4}<<&\frac{\lambda A_\mathrm{i}}{m_\phi}
\end{array}\label{typesa}
\end{align} 
in the annihilation case.

\subsection{\boldmath$\phi$ Damping and $\chi$ Heating Rates (Dynamics)}\label{ss:dynamics}
\subsubsection{Short Term}\label{sss:short term}
The damped energy per half-oscillation due to mediated damping found in Section~\ref{s:damping of phi in one half-oscillation} is summarized below

\begin{align}
\Delta \rho_{\psi\rightarrow\chi}\sim\left\{
\setlength\arraycolsep{1pt}
\begin{array}{LLRL}
\hfil|\dot{M}_\psi|^2\text{ } 
  &\text{for }&&T<<|\dot{M}_\psi|^{1/2} \\[2ex]
\hfil|\dot{M}_\psi| T^2\text{ }
   &\text{for }&|\dot{M}_\psi|^{1/2}<<&T<<\lambda A \\[2ex]
\hfil|\dot{M}_\psi|(\lambda A)^2\text{ }
   &\text{for }&\lambda A<<&T
\end{array}\right. \label{dampingrates}
\end{align} 
where
\begin{align}
|\dot{M}_\psi|\sim 
\begin{cases}
\displaystyle \hfil\lambda Am_\phi\min\left(1, \frac{\lambda A}{T} \right) & \text{\fcolorbox{black}{white}{\rule{0pt}{6pt}\rule{6pt}{0pt}} bare mass}\\[2ex]
\displaystyle \hfil\lambda^3A^2 & \text{\fcolorbox{black}{blue}{\rule{0pt}{6pt}\rule{6pt}{0pt}} thermal mass}\\[2ex]
\displaystyle \hfil\lambda T^2 & \text{\fcolorbox{black}{green}{\rule{0pt}{6pt}\rule{6pt}{0pt}} thermal log}\\[2ex]
\displaystyle \hfil\lambda^3A\max\left(\lambda^3A, T\right) & \text{\fcolorbox{black}{red}{\rule{0pt}{6pt}\rule{6pt}{0pt}} linear potential (annihilation)}
\end{cases}
\label{mpsidotda}
\end{align}
The conditions for and meaning of each regime are defined in Section~\ref{s:effective potential of phi}. It can be shown case by case that $A$ does not change significantly in one half-oscillation. Therefore, the damping process can be treated as continuous in the timescale of many oscillations of $\phi$.

\subsubsection{Long Term}\label{sss:long term}

\begin{table*}[]
\caption{Mediated damping and heating rates in the decay and annihilation case. The damping rates $\Gamma_\phi\equiv-\dot{A}/A$ tell us how fast the amplitude $A$ of $\phi$ decreases when the temperature $T$ is approximately constant and the heating rates $\text{L}_\chi\equiv\dot{T}/T$ tell us how fast the temperature $T$ of the thermal bath $\chi$ increases.\bigskip} 
\ra{2.2}
\setlength{\tabcolsep}{5pt}
\begin{adjustbox}{max width=\textwidth}
\begin{tabular}{l|>{\centering}m{1.7cm}|>{\centering}m{1.7cm}|>{\centering}m{1.7cm}|>{\centering}m{1.7cm}|>{\centering}m{1.7cm}|>{\centering\arraybackslash}m{1.7cm}|}
\cline{2-7}
\multirow{2}{*}{} & \multicolumn{2}{c|}{\centering $T<<|\dot{M}_\psi|^{1/2}$} & \multicolumn{2}{c|}{\centering $|\dot{M}_\psi|^{1/2}<<T<<\lambda A$} & \multicolumn{2}{c|}{\centering $\lambda A<<T$} \\ \cline{2-7} & $\Gamma_\phi$ & $\text{L}_\chi$ & $\Gamma_\phi$ & $\text{L}_\chi$ & $\Gamma_\phi$ & $ \text{L}_\chi$ \\ \hline
\multicolumn{1}{|l|}{\fcolorbox{black}{white}{\rule{0pt}{6pt}\rule{6pt}{0pt}} bare mass} & $\lambda^2m_\phi$ & $\displaystyle \frac{\lambda^2m_\phi^3A^2}{g_*T^4}$ & $\displaystyle \frac{\lambda T^2}{A}$ & $\displaystyle \frac{\lambda m_\phi^2 A}{g_*T^2}$ & $\displaystyle \frac{\lambda^4A^2}{T}$ & $\displaystyle \frac{\lambda^4m_\phi^2A^4}{g_*T^5}$ \\
\multicolumn{1}{|l|}{\fcolorbox{black}{blue}{\rule{0pt}{6pt}\rule{6pt}{0pt}}  thermal mass} & - & - & - & - & $\displaystyle \frac{\lambda^4A^2}{T}$ &  $\displaystyle \frac{\lambda^6A^4}{g_*T^3}$ \\
\multicolumn{1}{|l|}{\fcolorbox{black}{green}{\rule{0pt}{6pt}\rule{6pt}{0pt}} thermal log} & - & - & $\displaystyle\frac{\lambda T^2}{A}$ & $\displaystyle \frac{\lambda T^2}{g_*A}$ & - & - \\
\multicolumn{1}{|l|}{\fcolorbox{black}{red}{\rule{0pt}{6pt}\rule{6pt}{0pt}} linear potential (annihilation)} & $\lambda^7A$ & $\displaystyle \frac{\lambda^{17}A^5}{g_*T^4}$ & $\displaystyle \frac{\lambda T^2}{A}$ & $\displaystyle \frac{\lambda^5A}{g_*}$ & - & - \\[1.5ex] \hline
\end{tabular}
\end{adjustbox}
\label{table}
\end{table*}

In the timescale of many oscillations of $\phi$, the mediated damping rate of $\phi$ can be defined as
\begin{align}
\Gamma_{\phi}\equiv-\frac{\dot{A}}{A}\sim \frac{1}{\left<\dot{\phi}^2\right>} \frac{\Delta \rho_{\psi\rightarrow\chi}}{t_{\text{osc}}}
\end{align}
Using $M_\psi^2\sim \lambda^2\phi^2+T^2$, this becomes
\begin{align}
\Gamma_{\phi}\sim \left[\frac{\lambda}{|\dot{M}_\psi|}\min\left(1,\frac{\lambda A}{T}\right)\right]^2\frac{\Delta \rho_{\psi\rightarrow\chi}}{t_{\text{osc}}} \label{Gammaphi}
\end{align}
where $\Delta \rho_{\psi\rightarrow\chi}$ is given by Eq.~\eqref{dampingrates}, $|\dot{M}_\psi|$ is given by~\eqref{mpsidotda}, and
\begin{align}
t_{\text{osc}}\sim 
\begin{cases}
\displaystyle \hfil m_\phi^{-1} & \text{\fcolorbox{black}{white}{\rule{0pt}{6pt}\rule{6pt}{0pt}} bare mass}\\[2ex]
\displaystyle \hfil \frac{1}{\lambda T} & \text{\fcolorbox{black}{blue}{\rule{0pt}{6pt}\rule{6pt}{0pt}} thermal mass}\\[2ex]
\displaystyle \hfil\frac{A}{ T^2} & \text{\fcolorbox{black}{green}{\rule{0pt}{6pt}\rule{6pt}{0pt}} thermal log}\\[2ex]
\displaystyle \hfil\min\left(\frac{1}{\lambda^5A},\frac{1}{\lambda^2 T}\right) & \text{\fcolorbox{black}{red}{\rule{0pt}{6pt}\rule{6pt}{0pt}} linear potential (annihilation)}
\end{cases}
\label{toscda}
\end{align}
is the oscillation period of $\phi$\footnote{There appears to be a discontinuity when we go from $t_{\text{osc}}\sim m_\phi^{-1}$ in the bare mass regime, or $t_{\text{osc}}\sim\frac{1}{\lambda^2 T}$ in the thermal linear trapping regime, to $t_{\text{osc}}\sim \frac{1}{\lambda T}$ in the thermal mass regime. The actual transition is continuous, since the period of oscillation in the bare mass and thermal linear trapping regime can be written more precisely as $t_{\text{osc}}\sim m_\phi^{-1}\left(1-\frac{T}{\lambda A}\right)+\frac{1}{\lambda T}$ and $t_{\text{osc}}\sim\frac{1}{\lambda^2T}\sqrt{1-\frac{T}{\lambda A}}+\frac{1}{\lambda T}$ respectively.}. In Eq.~\eqref{Gammaphi}, the rate of energy being damped due to mediated damping is defined relative to $\left<\dot{\phi}^2\right>$ because it represents the part of the energy of $\phi$-$\psi$ system that depends on $A$. Putting the pieces together, the damping rate $\Gamma_\phi$ of $\phi$ in each of the regions is presented in Table~\ref{table}. We have also included in the table the thermal bath heating rate 
\begin{align}
\text{L}_\chi\equiv\frac{\dot{T}}{T}\sim\frac{1}{g_* T^4} \frac{\Delta \rho_{\psi\rightarrow\chi}}{t_{\text{osc}}} \label{Lchi}
\end{align} 

Apart from mediated damping, another significant source of $\phi$ damping is scatterings with $\psi$ particles. The damping rate due to scattering with $\psi$ particles is estimated in Appendix \ref{s:psiphipsiphi scattering} to be
\begin{align}
\Gamma_{\phi,\text{scat}}\sim \begin{cases}
\displaystyle  \hfil \frac{\lambda^3}{A}\max\left(|\dot{M}_\psi|,T^2\right) &\text{for }T<<\lambda A\\[2ex]
\displaystyle \hfil\lambda^4 T &\text{for }T>>\lambda A
\end{cases}
\label{scatdamp}
\end{align}
By comparing the above damping rate with the previously obtained mediated damping rate, we find that the mediated damping is dominant as long as
\begin{align}
A\gtrsim  T \label{scatteringdominates}
\end{align}
When the above condition is violated, the $\phi$ condensate will break into $\phi$ particles due to $\phi$-$\psi$ scatterings and subsequently thermalize. Once the fluctuation of $\phi$ grows to a significant level, the parameter space diagram (Figure~\ref{figure:summary}) will fail to give an adequate description of the system since it does not provide any inhomogeneity information of $\phi$.

\subsection{Full Damping Process}\label{ss:damping time}
Having understood the kinematics and dynamics of the system, we can proceed to study the full damping process. The damping process is not a simple exponential decay, and so we cannot assign a single timescale to the whole process. Nevertheless, having derived the $\phi$ damping and $\chi$ heating rates in every region in the parameter space means that we know the evolution rate ($\phi$ damping or $\chi$ heating rate) at every point along a trajectory, and the duration of the process is simply given by the point on the trajectory where the evolution rate is the slowest.

If the system starts with a large $A$ and an extremely small $T$, the system will in general go through three stages.

\subsubsection{\boldmath $\chi$ Heating Stage}\label{sss:heating stage}
Initially, when the energy of the $\phi$ oscillation is much greater than that of the thermal bath, the timescale of evolution of the system is given by the heating rate of the thermal bath $\text{L}_\chi$, defined in Eq.\eqref{Lchi}. According to the parameter space diagram, Figure~\ref{figure:summary}, the heating rate $\text{L}_\chi$ up to the point where $T$ stops increasing significantly is always decreasing along the trajectory. Therefore, the time $t_{\text{heat}}$ it takes for the thermal bath $\chi$ to heat up to close to its maximum temperature is determined by the heating rate at the end of the heating process where $T\simeq T_{\max}$, which can be found with the aid of Table~\ref{table} to be
\begin{align}
t_{\text{heat}}\sim \left.\text{L}_\chi^{-1}\right|_{\rho_\phi+\rho_\psi\sim\rho_\chi}
\sim \left\{
\setlength\arraycolsep{1pt}
\begin{array}{CL}
\frac{ m_\phi^{1/2}}{g_*^{1/4}\lambda^4A_\mathrm{i}^{3/2}}    &\text{ Type I }\\[2ex]
\frac{g_*^{1/2} }{\lambda m_\phi} &\text{ Type II }\\[2ex] 
\frac{g_*^{3/4}}{\lambda^3 m_\phi^{1/2}A_\mathrm{i}^{1/2}} &\text{ Type III }\\[2ex]
\frac{g_*^{3/4}}{\lambda^{11/2} A_\mathrm{i}} &\text{ Type IV }
\end{array}\right. \label{theat}
\end{align} 
where $A_\mathrm{i}$ is the initial amplitude of $\phi$ when $T$ is extremely small and the condition for each trajectory type is given by Eqs.~\eqref{typesd} and \eqref{typesa}.

\subsubsection{\boldmath $\phi$ Damping Stage}\label{sss:damping stage}
After the temperature of $\chi$ stops growing significantly, the amplitude of $\phi$ will decrease to exponentially smaller values due to mediated damping. Since, according to Figure~\ref{figure:summary}, there is no more local minimum of the evolution rate after $T\simeq T_{\max}$, the total time $t_{\text{damp}}(A_\mathrm{f})$ it takes for the amplitude of $\phi$ to be damped to a final amplitude $A_\mathrm{f}<<A_\mathrm{i}$ is given by either the damping rate at the beginning or at the end of the damping
\begin{align}
t_{\text{damp}}(A_\mathrm{f})&\sim\max\left(t_{\text{heat}}, \left.\Gamma_\phi^{-1}\right|_{A=A_\mathrm{f}}\right)\sim\begin{cases} 
\displaystyle \hfil t_{\text{heat}} &\text{for }\displaystyle A_\mathrm{f}\gtrsim\frac{T_{\max}}{\lambda}\\[2ex]
\displaystyle \hfil\max\left(t_{\text{heat}},\frac{ T_{\max}}{\lambda^4A_\mathrm{f}^2}\right) &\text{for }\displaystyle A_\mathrm{f}\lesssim\frac{T_{\max}}{\lambda}
\end{cases}
\label{dampinguntilAf}
\end{align}
where $t_{\text{heat}}$ can be found in Eq.~\eqref{theat} and $T_{\max}$ can be found in Eq.~\eqref{Tmaxdecay} and \eqref{Tmaxannihilation}. As we can see, depending on how small $A_\mathrm{f}$ is, it may take less or more time than $t_{\text{heat}}$ for the amplitude of $\phi$ to decrease substantially. The total damping time until thermalization of $\phi$ begins at $A_\mathrm{f}\sim T_{\max}$ is
\begin{align}
t_{\text{damp}}\left(A_\mathrm{f}\sim T_{\max}\right)&\sim \max\left(t_{\text{heat}}, \frac{1}{\lambda^4 T_{\max}}\right)\sim \frac{1}{\lambda^4 T_{\max}}
 \label{dampingtime}
\end{align}
Where in the last step we assumed that $g_*\lesssim 10$, in which case the time it takes for $\phi$ to be damped until thermalization of $\phi$ begins is always longer than or of the same order with the heating time $t_{\text{heat}}$ of the thermal bath.

\subsubsection{\boldmath $\phi$ Thermalization Stage}\label{sss:preheating and thermalization}
After the amplitude of $\phi$ is damped to $A\lesssim T_{\max}$, the system enters a regime where $\phi$-$\psi$ scatterings dominate the damping effect and the $\phi$ field thermalizes. Since the damping due to $\phi$-$\psi$ scatterings, Eq.~\eqref{scatdamp}, in the $T>>\lambda A$ regime is independent of $A$ the damping process in this regime is purely exponential. Consequently, $\phi$ can be damped completely and, at the same time, thermalized. The thermalization timescale of $\phi$ is simply given by the inverse of the damping rate due to scatterings
\begin{align}
\tau_{\text{th}}=\Gamma_{\phi,\text{scat}}^{-1}\sim \frac{1}{\lambda^4 T_{\max}} \label{phthtime}
\end{align}
which should naturally be of the same order as the time $t_{\text{damp}}\left(A_\mathrm{f}\sim T_{\max}\right)$ it takes for $A$
to reduce to $T_{\max}$ since the damping rate does not change in the scattering regime. Thus, for $g_*\lesssim 10$, $t_{\text{damp}}\left(A_\mathrm{f}\sim T_{\max}\right)\sim\tau_{\text{th}}$ is the longest timescale of the system. Interestingly, it is shorter for systems with higher initial amplitudes $A_\mathrm{i}$, because they eventually lead to higher $T_{\max}$.

\section{Summary}\label{s:summary}
\begin{figure}[H]
\centering
\includegraphics[scale=0.25]{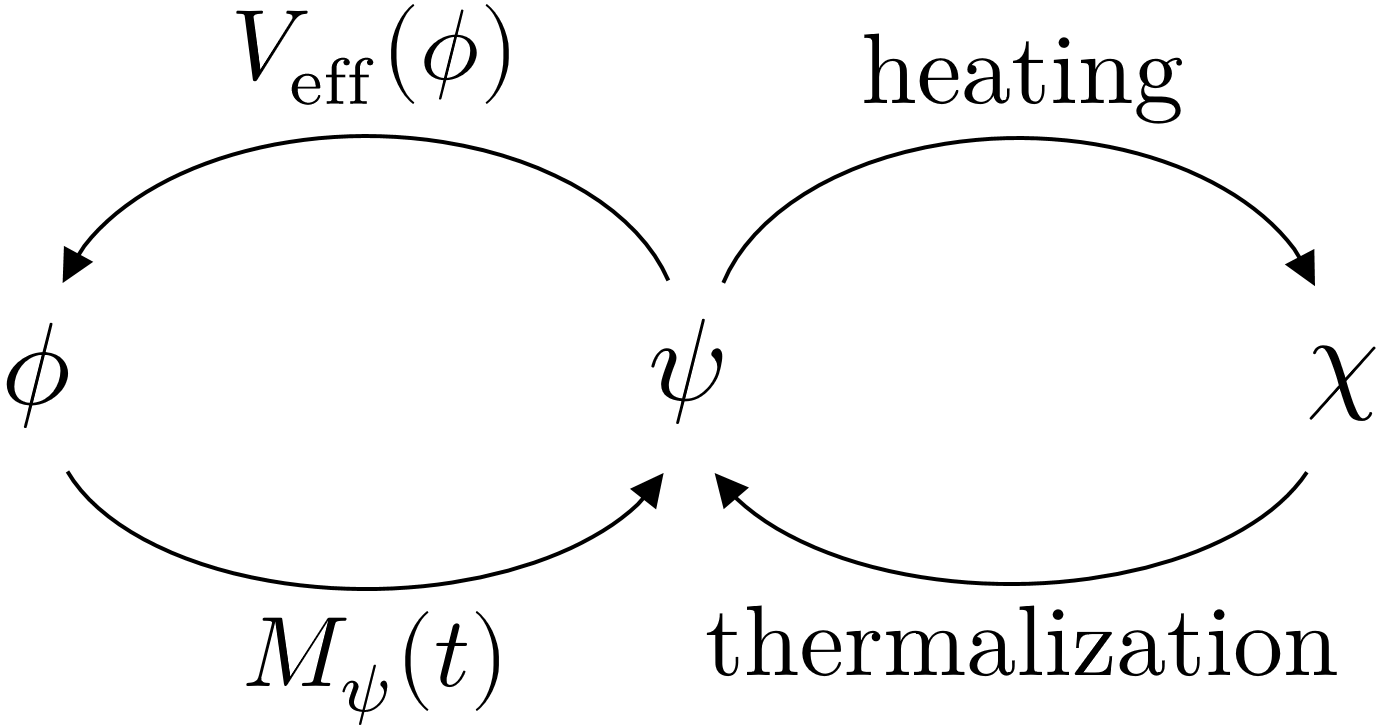}
\caption{Processes involved in the damping of $\phi$.}
\label{figure:method}
\end{figure}
The processes involved in the damping of $\phi$ are summarized in Figure \ref{figure:method}. All transfer of energy in the system involves $\psi$. Hence, the mediated damping rate is determined once we know the dynamics of $\psi$, which is driven by the time-dependent effective mass $M_\psi(t)$ it receives from the oscillating $\phi$ field and the thermalizing effect of the thermal bath $\chi$. Treating the $\phi$ field as a background, we evaluated the dynamics of $\psi$ using the integrated Boltzmann Eqs.~\eqref{boltzmannd} and \eqref{boltzmanna}, taking into account possible non-adiabatic $\psi$-particle production in an ad-hoc manner. We also took into account the complication that when $\psi$ gets highly excited it may significantly affect $\phi$ via the effective potential $V_{\text{eff}}(\phi)$ underlying the $\phi$ oscillation. The different types of $V_{\text{eff}}(\phi)$ and their regimes of applicability are studied in Section~\ref{s:effective potential of phi}.

Our main results are best summarized by Figure~\ref{figure:summary} (kinematics) and Table~\ref{table} (dynamics). Figure~\ref{figure:summary} shows the parameter space of our system, which is divided into various regions based on the form of $V_{\text{eff}}(\phi)$ and the nature of the $\psi$ dynamics. In both decay and annihilation cases, the nature of the $\psi$ dynamics was found to be determined by three scales: a scale $|\dot{M}_\psi|^{1/2}$ representing the rate of change of $\psi$'s effective mass, the temperature $T$ of $\chi$, and $\lambda A$ which determines the amplitude of the mass variation of $\psi$ due to its interaction with $\phi$. There are three cases depending on the size of $T$ relative to the other two scales:
\begin{itemize}
\item Low temperature regime, $T<<|\dot{M}_\psi|^{1/2}$\\
Thermal effects are negligible, and thus the situation reduces to that of instant preheating. Typical evolution of the number density $n_\psi$ of $\psi$ in this case during a half-oscillation of $\phi$ is shown in Figures \ref{figure:lowdecay} and \ref{figure:lowannihilation}.
\item Intermediate temperature regime, $|\dot{M}_\psi|^{1/2}<<T<<\lambda A$\\
Thermal effects dominate over non-adiabatic effects but are limited to short intervals where the mass of $\psi$ is small enough to avoid Boltzmann suppression. Typical evolution of the number density $n_\psi$ of $\psi$ in this case during a half-oscillation of $\phi$ is shown in Figures \ref{figure:intermediatedecay} and \ref{figure:intermediateannihilation}.
\item High temperature regime, $\lambda A<<T$\\
The $\psi$ field is always close to thermal equilibrium without Boltzmann suppression. Typical evolution of the number density $n_\psi$ of $\psi$ in this case during a half-oscillation of $\phi$ is shown in Figure~\ref{figure:high}.
\end{itemize}
Once the dynamics of $\psi$ is known, the amount of mediated damping per half-oscillation of $\phi$ can be calculated using Eq.~\eqref{temperature}. In the timescale of many oscillations of $\phi$, the mediated damping rates of $\phi$ and the resulting heating rates of $\chi$ can be found using Eqs.~\eqref{Gammaphi} and \eqref{Lchi}. The results are summarized in Table~\ref{table}.

Knowing the damping rate in each regime in the parameter space, we can study the long term dynamics of the system. If initially the temperature of the thermal bath is extremely low, the system will go through three stages:
\begin{enumerate}
\item $\chi$ heating stage\\
Initially, mediated damping barely needs to decrease the amplitude $A$ to increase the temperature $T$. The timescale of the process at this stage is given by the heating rate $\text{L}_\chi\equiv \dot{T}/T$ of the thermal bath. The temperature of $\chi$ will continue to increase until it approaches its maximum value, $T_{\max}$, given in Eqs.~\eqref{Tmaxdecay} and \eqref{Tmaxannihilation}. This stage lasts for $t_{\text{heat}}$, which was found in Eq.~\eqref{theat}.
\item $\phi$ damping stage\\
After the temperature $T$ essentially stops increasing, the amplitude $A$ will decrease to exponentially smaller values. The timescale of the process at this stage is given by the damping rate $\Gamma_\phi\equiv -\dot{A}/A$ of $\phi$. It takes around $t_{\text{damp}}(A_\mathrm{f})$, found in Eq.~\eqref{dampinguntilAf}, for $A$ to reduce to $A_\mathrm{f}$.
\item $\phi$ thermalization stage\\
Once $A$ reduces to $A\sim T_{\max}$, scattering effects, as calculated in Appendix \ref{s:psiphipsiphi scattering}, take over as the dominant damping effect, and the $\phi$ field begins to thermalize. The time it takes for the $\phi$ field to thermalize is determined by the inverse of the damping rate due to scattering $\Gamma_{\phi,\text{scat}}$, Eq.~\eqref{phthtime}.
\end{enumerate}
In general, the system may start in any of the three stages above and go through the subsequent stages. The time it takes for the system to evolve through any segment of the parameter space trajectory can be calculated by identifying the point of maximum half-time (inverse of damping or heating rate) of the segment of interest with the aid of Figure~\ref{figure:summary} and finding the appropriate half-time in Table~\ref{table}.

\section{Discussion}\label{s:discussion}
\subsection{Comparison with Previous Works}\label{ss:comparison with previous works}

Our work is closely related to that of Refs. \citep{Mukaida2012, Mukaida2013, Moroi2013}. Refs. \citep{Mukaida2012, Mukaida2013} investigate the damping process of the oscillating $\phi$ field in essentially the same system as the present one using the two-particle irreducible (2PI) formalism. There are minor differences between their models and ours, such as our mediator field is real while theirs is complex, we consider the change in the temperature of the thermal bath due to energy transfer from the mediator field while they do not, while we limit our exploration to $\kappa \sim 1$ regime of the $\psi$-$\chi$ effective coupling they explore some of the $\kappa<<1$ regime, and Ref. \citep{Mukaida2012} also considers a fermionic mediator field. We reproduced their results and extend the analysis to the case where the mediator field particles are individually stable but pair annihilate. Ref. \citep{Moroi2013} considers a similar configuration with the stable mediator field case taken into account, but it focuses on the trapping instead of damping effect of the $\phi$ field and the $\phi$ field in their model has a symmetry breaking potential.

\subsubsection{Damping Effects}\label{ss:damping effects}
As in Refs. \citep{Mukaida2012, Mukaida2013}, one can calculate the damping rate of $\phi$ using the 2PI formalism \citep{jordan,berges,calzetta, calzetta2}, which, apart from the 2PI truncation, is exact. However, using such an elaborate approach when it is not absolutely needed may obscure the physics. Instead, having an understanding of a phenomenon at an elementary level is important as it makes it easier to spot similar phenomena in different circumstances.

In the 2PI formalism, the equation of motion of $\phi$ can be written as
\begin{align}
\ddot{\phi}(t)+\left[m_\phi^2+\Pi^{\text{(local)}}_{k=0}(t)\right]\phi(t)+\int_{t_{\text{i}}}^{t}dt' \Pi^{-}_{k=0}(t,t')\phi(t')=0 \label{2pi}
\end{align}
where all the interaction effects are contained in the self-energies $\Pi_{k}$ of $\phi$, with $\Pi^{\text{(local)}}_{k=0}(t)$ being the local part and $\Pi^{-}_{k}$ being the antisymmetric part of the non-local part. Refs. \citep{Mukaida2012, Mukaida2013} consider three types of damping coming from the leading local effect, leading non-local effect, and non-local effects at higher loop orders. It is useful to understand how the 2PI formalism corresponds to the language we use here. 

First, the effect of the local part of the self-energy of $\phi$ can be viewed as an extra time-dependent potential felt by $\phi$, whose asymmetry leads to damping that we have been referring to as mediated damping. In \citep{Mukaida2013}, the damping effect due to the local part of the self-energy is interpreted as an effective three-point decay process $\phi\rightarrow\psi\psi$ due to the resemblance of the damping rate to that of the perturbative decay due to the three scalar field interaction $\Lambda\phi\psi^2=\lambda^2\left<\phi\right>\phi\psi^2$. In the common cases that we consider, our results completely agree.

Second, if the time-correlation length of the non-local part of the self-energy of $\phi$ is much shorter than the oscillation period of $\phi$, the non-local integral in the 2PI equation of motion can be approximated with a local dissipative term that is proportional to $\dot{\phi}$ \citep{Mukaida2012,Mukaida2013,Moroi2013}. The term yields a friction-like damping of $\phi$ that is always present throughout the motion of $\phi$. In the particle picture, this corresponds to damping due to direct scattering by the particles of the mediator field, which we have estimated in Appendix \ref{s:psiphipsiphi scattering} to be given by Eq.~\eqref{scatdamp}, in agreement with the results obtained in \citep{Mukaida2013} using the 2PI formalism. 

Third, when $\psi$ particles are scarce, $\phi$ can still interact with $\chi$ through higher loop effects. This interaction results in an extra $\phi$ damping given by \citep{Mukaida2012,Mukaida2013} 
\begin{align}
\Gamma_{\phi, \text{higher-loop}}\sim \frac{b\lambda T^2}{A} \label{dim5}
\end{align}
where typically $b\sim 10^{-3}$. We included in our analysis the higher loop effects on the potential of $\phi$, i.e.\ the thermal log potential, but not its inherent damping effect. Neglecting the damping effect is justified since, as found in \citep{Mukaida2013}, the contribution from this effect is always insignificant compared to other contributions when we consider a full half-oscillation.

\subsubsection{Long Term Dynamics}\label{sss:long term damping process}
In assessing the long term dynamics of the system, the approach of Refs. \citep{Mukaida2012} is the complement of ours. They start by assuming that $\Gamma_\phi<<H$, and then evolve the relevant quantities, considering only the effects of the expansion of the universe until $\Gamma_\phi=H$, at which point they say that $\phi$ evaporates. We, on the other hand, neglect the Hubble expansion in our calculations and consider the internal dynamics of the system, following the evolution of the system from the point where the amplitude of the $\phi$ field is extremely large to the point where it thermalizes. Since the internal dynamics of the system involves many timescales, i.e.\ the thermal bath heating time $t_{\text{heat}}$, the $\phi$ damping time $t_{\text{damp}}$, and the $\phi$ thermalization time $\tau_{\text{th}}$, there will be cases that involve a combination of both Hubble expansion scaling and internal dynamics of the system. In those cases, neither of the approaches is sufficient on its own but our parameter space diagram (Figure~\ref{figure:summary}) will still be useful though the trajectories will be more complicated due to the Hubble expansion.

\subsection{Inhomogeneity of $\phi$}\label{ss:inhomogeneity of the phi field}
In the main analysis, we neglected the inhomogeneity in the $\phi$ field. If $\phi$-$\psi$ scattering is the only significant source of $\phi$ fluctuation, this assumption will be justified as long as Eq.~\eqref{scatteringdominates} is satisfied. Here we discuss more subtle and difficult to estimate effects that might amplify inhomogeneities in $\phi$. If these effects are strong, the inhomogeneous modes of $\phi$ may grow to significant levels earlier than the $\phi$-$\psi$ scattering calculation tells us. It is difficult to analytically deduce what will happen in those cases, but the present homogeneous analysis may serve as a first step towards understanding the full dynamics.

There are several mechanisms that may cause inhomogeneities in $\phi$ to grow. In the thermal log and linear potential regimes, they may grow due to the form of the effective potential $V_{\text{eff}}(\phi)$. For instance, this could happen when the effective mass of $\phi$ becomes negative or changes non-adiabatically. Also, when $\psi$ freezes out in the linear potential regime, any inhomogeneities in $\phi$ will lead to a force on $\psi$ particles pushing them to regions with smaller $\phi$, and the resulting inhomogeneity in $\psi$ would backreact to $\phi$, tending to further increase its inhomogeneity. Further, the amplitude dependence of the $\phi$ damping rate $\Gamma_\phi(A)$ may either amplify ($\partial\Gamma_\phi/\partial A<0$) or damp ($\partial\Gamma_\phi/\partial A>0$) inhomogeneities in $\phi$.

The $\phi$ field may also disintegrate into long-living non-topological solitons if its potential is flatter than quadratic \citep{Kasuya2002,Copeland:1995fq}. The tendency to disintegrate is because flatter than pure quadratic potential implies an attractive force between particles, meaning that it is energetically favourable for the particles to clump together to form clusters.

\subsection{Supersymmetric Flat Direction Damping}\label{ss:realistic supersymmetric flat direction damping}
We have considered a simple toy model consisting of an oscillating real scalar field $\phi$, a real scalar field $\psi$ acting as a mediator, and a thermal bath $\chi$. However, in realistic supersymmetric models the fields involved are necessarily complex. Nevertheless, whether $\psi$ and $\chi$ field are complex does not have any effect on the process beyond the numbers of degrees of freedom and, if the field-space motion of $\phi$ is nearly radial, the problem can be reduced to that of a real $\phi$ field.

In cases where the $\phi$ field has significant angular momentum, its complex treatment is inevitable. Let us speculate how the $\phi$ damping process in the complex field case changes compared to the real field case. For the mediated damping process to work efficiently, the value of $|\phi|$ has to dip low enough to allow the $\psi$ fields to be excited. One distinctive feature that makes the complex $\phi$ case different from the real one is that a complex field typically does not pass close to the origin of its field space where $\psi$ can be excited. Even if the angular momentum of the $\phi$ field is not conserved and nothing else prevents the field from passing arbitrarily close to the origin, passing close to the origin is still a rare event because of the smallness of the field space regime it occupies. This may prevent an otherwise strong mediated damping from taking place.

Most of the time, $\phi$ would wander far from the origin of the field space, causing $\psi$ to be heavy and difficult to excite. Occasionally, $\phi$ would pass close to the origin of the complex field space where the mass of $\psi$ is small, allowing significant mediated damping to take place. To simplify the problem, we can divide the motion of $\phi$ into separate swings or passings, each with duration of order $m_\phi^{-1}$, amplitude $|\phi|_{\max}$ of order $A$, and a field-space impact parameter $|\phi|_{\min}$ which takes on a random value between zero and $A$. Thus, in this view, there is an $a/A$ chance per passing for the impact parameter to be smaller than $a$.

The field space trajectory of $\phi$ allows many possible impact parameters to be sampled in a timescale much longer than the oscillation period of $\phi$. On average, we expect the occurrence of significant mediated damping in the complex $\phi$ case to be suppressed relative to the real $\phi$ case by a factor representing the probability for $\phi$ to swing close enough to the origin for $\psi$ to be excited
\begin{align}
\min\left(\frac{\max\left(|\dot{M}_\psi|^{1/2},T\right)}{\lambda A},1\right)
\end{align}
where $\max\left(|\dot{M}_\psi|^{1/2},T\right)/\lambda$ is the size of the regime in the field space of $\phi$ within which $\psi$ can be excited. As a crude approximation, we can slightly modify the results for the real $\phi$ case to accommodate the complex $\phi$ case by multiplying the $\phi$ damping and $\chi$ heating rates we previously obtained with the field space suppression factor
\begin{align}
\Gamma_{\phi}&\rightarrow \Gamma_{\phi}\times \min\left(\frac{\max\left(|\dot{M}_\psi|^{1/2},T\right)}{\lambda A},1\right)\\
\text{L}_{\chi}&\rightarrow \text{L}_{\chi}\times \min\left(\frac{\max\left(|\dot{M}_\psi|^{1/2},T\right)}{\lambda A},1\right)
\label{suppresseddamping}
\end{align}
After the factor is included, the parameter space diagram in the complex $\phi$ case is still given by Figure~\ref{figure:summary}, except that the local minima of the evolution rate for type III trajectories are moved to the intersections of the trajectories with the linear potential regime boundary and those for type IV trajectories are moved to the intersections of the trajectories with the $T=|\dot{M}_\psi|^{1/2}$ line. The thermal bath heating time then becomes
\begin{align}
t_{\text{heat}}\sim \left.\text{L}_{\chi}^{-1}\right|_{\rho_\phi+\rho_\psi\sim\rho_\chi}\sim \left\{
\setlength\arraycolsep{1pt}
\begin{array}{CL}
\frac{ m_\phi^{1/2}}{g_*^{1/4}\lambda^4A_\mathrm{i}^{3/2}}    &\text{ Type I }\\[2ex]
\frac{g_*^{3/4} A_\mathrm{i}^{1/2}}{m_\phi^{3/2}} &\text{ Type II }\\[2ex] 
\frac{g_* }{\lambda^2 m_\phi} &\text{ Type III }\\[2ex]
\frac{g_* }{\lambda^{7} A_\mathrm{i}} &\text{ Type IV } \end{array}\right. \label{theatsuppressed}
\end{align}
where the condition for each trajectory type is given in Eqs.~\eqref{typesd} and \eqref{typesa}. Furthermore, Eq.~\eqref{dampinguntilAf} remains valid since it merely relies on the non-existence of local minima of the evolution rate after $T\simeq T_{\max}$. If we assume $g_*\sim 10$, $t_{\text{damp}}(A_\mathrm{f}\sim T_{\max})$ is now only longer than $t_{\text{heat}}$ for type I trajectories, unlike in the real $\phi$ case. The damping time is thus given by
\begin{align}
t_{\text{damp}}(A_\mathrm{f}\sim T_{\max})&\sim \max\left(\frac{1}{\lambda^4 T_{\max}},t_{\text{heat}}\right)\sim 
\begin{cases}
\displaystyle \hfil\frac{1}{\lambda^4 T_{\max}} &\text{Type I}\\[2ex]
\displaystyle \hfil t_{\text{heat}} &\text{Type II, III, IV}
\end{cases}
\end{align}
i.e.\ for large enough initial amplitudes of $\phi$, it would take more time for the thermal bath to heat up than for $\phi$ to thermalize.

The crude estimates above may be useful in the context of low energy scale Affleck-Dine (AD) leptogenesis \citep{Stewart1996,Jeong:2004hy,Kim:2008yu} taking place after thermal inflation \citep{Lyth1995,Lyth1996}. In the usual, high energy AD baryogenesis \citep{AD1985,Dine:1995kz}, the job of damping the AD fields to preserve the asymmetry is done by Hubble friction. However, in the low energy scenario, the Hubble friction is negligible and instead preheating among the AD fields (corresponding to $\phi$ in our setup) provides sufficient initial damping \citep{Kawasaki2006,Felder2007,Kim:2008yu}, but the effects of the mediator fields $\psi$ (squarks plus some others) and the thermal bath $\chi$ (gluons plus some others), may be needed to sustain the damping at later times.

There is more complication in the complex field case. Apart from fluctuations in the amplitude of the $\phi$ field, which are discussed in Section~\ref{ss:inhomogeneity of the phi field}, fluctuations may also grow in the angular/phase direction of the field. In some cases, stable localized configurations called Q-balls \citep{Coleman1985}, whose stability relies on the conservation of a $U(1)$ charge may form. Even if the initial condition has zero charge, Q-balls and their negative counter parts, anti Q-balls, could be abundantly pair-produced. Q-ball formation may impede the mediated damping process because it suppresses radial oscillations of the $\phi$ field. The fate of the $\phi$ field will depend highly on how rapidly the Q-balls form compared to its damping rate. Q-ball formation in the context of a flat-direction oscillating under a supersymmetry-breaking potential \citep{Kusenko1997b,Kusenko1997,Kasuya:1999wu,Kasuya:2000wx,Kasuya2002,Enqvist:1999mv} and a thermal logarithmic potential \citep{Kasuya:2010vq, Chiba:2010ff} have been studied extensively, but it has not been done specifically in the present context. We leave the analysis of Q-ball formation in the present context to future publications.

\section{Conclusion}\label{s:conclusion}
We have studied a toy model where a homogeneous oscillating real scalar field $\phi$ is coupled to a thermal bath $\chi$ indirectly through a mediator field $\psi$. The model is useful for understanding the damping process of a supersymmetric flat direction condensate in the early universe. We reproduced the results of Refs. \citep{Mukaida2012, Mukaida2013} in the case where the mediator field particles can decay to the thermal bath using more direct, elementary methods and extended the study to the case where the mediator field particles are individually stable but pair annihilate. In the latter case, the mediator field may freeze out and hence give different varieties of effective potential to the oscillating field, making the damping process more complicated.

If initially the temperature of the thermal bath is extremely low, the system will go through three stages. In the beginning, mediated damping barely needs to reduce the amplitude of the oscillating field to heat up the thermal bath, and so we describe this stage in terms of the heating rate of the thermal bath. After about half of the energy of the system has been transferred to the thermal bath, the temperature of the thermal bath stops increasing significantly while the amplitude of the oscillating field decreases to exponentially smaller values, and so we describe this stage in terms of the damping rate of the oscillating field. Finally, the oscillating field begins to thermalize when its amplitude has reduced to the order of the thermal bath temperature. In general, the system may start in any of the three stages and go through the subsequent stages.

After mapping the parameter space of the system (Figure~\ref{figure:summary}) based on the nature of the effective potential of the oscillating field and the dynamics of the mediator field, we calculated the damping rates of the oscillating field and the resulting heating rates of the thermal bath in each regime, the results of which are summarized in Table~\ref{table}. We then used the results to study the full dynamics of the system. Once the damping process begins, it can get stronger or weaker depending on the nature of the damping and how the parameters that affect the damping change. As a result, the process will have different timescales at different stages of the process, meaning that we cannot associate the whole process with a single timescale. Instead, as we have done in Section~\ref{ss:damping time}, the timescale of the process has to be determined case by case depending on the initial and final state of interest.

Our results suggest that, for a given effective potential of the oscillating field, stable and unstable mediator fields give the same damping effect up to $O(1)$ factors. While unstable mediator fields tend to follow their equilibrium abundances, stable mediator fields may freeze out above their equilibrium abundances and consequently give steeper effective potentials to the oscillating field than the unstable ones. This implies that, if the oscillating scalar field couples to both stable and unstable mediator fields, the damping effect would resemble that due to the stable mediator fields, as they tend to dominate the effective potential. The distinction between stable and unstable mediator fields fades away when the amplitude of the oscillating field is small enough that the effective mass of the mediator field remains smaller than the temperature of the thermal bath.

As discussed in Section~\ref{s:discussion}, compared to realistic situations of supersymmetric flat direction damping, our work has a number of limitations, which deserve to be addressed in future work. Most important are the growth and effects of inhomogeneities, including the possible formation of non-topological solitons such as Q-balls. Also, flat directions in the MSSM, or extensions thereof, are multidimensional, as opposed to the single real field $\phi$ of our toy model, and couple to numerous other fields, as opposed to our single real field $\psi$, in a complex way, allowing rich dynamics within the $\phi$ and $\psi$ sectors, not just between them. However, we hope our work will provide some useful guidance in studying such challenging early universe physics.

\section*{Acknowledgements}
The authors thank Duong Q. Dinh, Wanil Park, and Hugo Serodio for helpful discussions. EHT is indebted to the hospitality of the Center for Axion and Precision Physics (CAPP) and especially to Jonghee Yoo. EHT is supported in part by the Institute for Basic Science (IBS-R017-D1-2017-a00/IBS-R017-G1-2017-a00).

\section*{Appendix}
\appendix

\section{Cross Sections}\label{s:cross section times relative velocity}
The thermally averaged cross-section $\sigma_{12\rightarrow34}$ times Lorentz-invariant relative velocity $v_{\text{rel}}=|\vec{v}_1-\vec{v}_2|$ for a scattering process of the form $1+2\rightarrow3+4$ can be written in terms of centre of mass frame variables of the scattering particles as follows \citep{Peskin1995}
\begin{align}
\big<\sigma_{12\rightarrow34}v_{\text{rel}}\big>=\frac{p_f|\mathcal{M}|^2}{16\pi E_{1}E_{2}\left(E_1+E_2\right)}
\end{align}
where $E_1$ and $E_2$ are the energies of the incoming particles, $p_f$ is the magnitude of the spatial momentum of any one of the scattering products, and $\mathcal{M}$ is the scattering amplitude.

\subsection{\boldmath $\psi\psi\rightarrow\chi\chi$}\label{s:psipsichichi}
Since $\psi$ is never ultra-relativistic, we have $E_1\sim E_2\sim M_\psi$. Furthermore $p_f\sim \sqrt{E_\psi^2-m_\chi^2}\sim E_\psi\sim M_\psi$ and $|\mathcal{M}|^2=g_*\kappa_{\text{A}}^4$. Therefore
\begin{align}
\big<\sigma_{\psi\psi\rightarrow\chi\chi}v_{\text{rel}}\big>\sim \frac{g_*\kappa_{\text{A}}^4}{ M_\psi^2}= \frac{\kappa }{M_\psi^2}\sim \frac{1}{M_\psi^2}
\end{align}

\subsection{\boldmath $\psi\phi\rightarrow\psi\phi$}\label{s:psiphipsipshi}
Assuming that both $\psi$ and $\phi$ are not ultrarelativistic, we have $E_1\equiv E_{\phi}\sim M_\phi$ and $E_2\equiv E_{\psi}\sim M_\psi$. The $\phi$ particles are initially at rest when $\phi$ is homogeneous and are much lighter than the $\psi$ particles, so, after each collision between a $\psi$ particle and a $\phi$ particle, the $\psi$ particle will barely change its speed and the $\phi$ particle will acquire a relative speed $v_{\text{rel}}$ of the order of the $\psi$ particle's speed, which is roughly $v_\psi\sim \max\left(|\dot{M}_\psi|^{1/2},T\right)/M_\psi$, where $\max\left(|\dot{M}_\psi|^{1/2},T\right)$ is the typical momentum of a $\psi$ particle. After the collision, since the $\phi$ particle is much lighter, it will contribute most of the relative velocity in the center of mass frame, i.e.\ $|v_\phi^\prime|\approx v_{\text{rel}}\sim \max\left(|\dot{M}_\psi|^{1/2},T\right)/M_\psi$. Therefore
\begin{align}
p_f=|p_\phi^\prime|\approx M_\phi v_{\text{rel}}=\max\left(|\dot{M}_\psi|^{1/2},T\right)\frac{M_\phi}{M_\psi}
\end{align}
Putting everything together, we get
\begin{align}
\left<\sigma_{\psi\phi\rightarrow\psi\phi}v_{\text{rel}}\right>\sim\frac{\lambda^4\max\left(|\dot{M}_\psi|^{1/2},T\right)}{M_\psi^3}
\end{align}

\section{Thermal Decoupling}\label{s:point of separation}
Assuming that the $\psi$ field decouples from the thermal bath $\chi$ when the $\psi$ particles are non-relativistic, we can apply the following approximation
\begin{align}
n^{\text{eq}}_\psi\sim M_\psi^{3/2}T^{3/2}e^{-M_\psi/T}
\end{align}
According to Eqs.~\eqref{stick} and \eqref{separation}, thermal decoupling occurs when 
\begin{align}
e^{-M_\psi/T}\sim \dfrac{|\dot{M_\psi}|M_\psi^{1/2}}{T^{3/2}}
\end{align}
Therefore, the number density of $\psi$ at the time of thermal decoupling $t_{\text{TD}}$ is 
\begin{align}
n_\psi(t_{\text{TD}})&\sim \left(1+\xi\right) T^3\left[\dfrac{M_\psi(t_{\text{TD}})}{T}\right]^{3/2}e^{-M_\psi(t_{\text{TD}})/T}\sim  |\dot{M}_\psi|T 
\end{align}
where we have used $M_\psi(t_{\text{TD}})\sim T$.

\section{\boldmath $\psi\phi\rightarrow\psi\phi$ Scattering}\label{s:psiphipsiphi scattering}
We can treat $\phi$ as a condensate of zero-momentum $\phi$-particles with number density \citep{towards}
\begin{align}
n_\phi=\frac{\frac{1}{2}M_\phi^2A^2}{M_\phi}=\frac{1}{2}M_\phi A^2
\end{align}
where $M_\phi$ is the effective mass of $\phi$. The rate of occurrence of the $\psi\phi\rightarrow\psi\phi$ process, or equivalently the decay rate of the number density corresponding to the zero-mode of $\phi$, is given by
\begin{align}
\dot{n}_{\text{scat}}\approx -n_\psi n_\phi\left<\sigma_{\psi\phi\rightarrow\psi\phi} v_{\text{rel}}\right>\sim -\frac{\lambda^4 n_\psi n_\phi \max\left(|\dot{M}_\psi|^{1/2},T\right)}{M_\psi^3}
\end{align}
where $\left<\sigma_{\psi\phi\rightarrow\psi\phi} v_{\text{rel}}\right>$ is calculated in Appendix \ref{s:cross section times relative velocity}. The oscillation-averaged damping rate due to $\phi\psi\rightarrow\phi\psi$ scattering is thus given by
\begin{align}
\Gamma_{\phi,\text{scat}}=-\frac{\dot{A}}{A}\sim-\frac{M_\phi\left<\dot{n}_{\text{scat}}\right>}{\left<\dot{\phi}^2\right>}\sim \begin{cases}
\displaystyle \hfil\lambda^4 \frac{\max\left(|\dot{M}_\psi|,T^2\right)}{\lambda A} &\text{for }T<<\lambda A\\[2ex]
\displaystyle \hfil\lambda^4 T &\text{for }T>>\lambda A
\end{cases}
\end{align}
where $\left<\hdots\right>$ denotes oscillation average. Note that the above result is expected to be valid only at timescales much longer than the oscillation period of $\phi$ where its particle interpretation is legitimate.

\bibliography{references}
\bibliographystyle{JHEP}

\end{document}